\documentclass[aps,prl,floatfix,superscriptaddress,twocolumn,footinbib]{revtex4-2}
\usepackage[T1]{fontenc}
\usepackage[latin9]{inputenc}
\usepackage{xcolor}
\usepackage{mathrsfs}
\usepackage{amsmath}
\usepackage{amssymb}
\usepackage{makeidx}
\usepackage{amsfonts,amssymb}
\makeindex
\usepackage{graphicx}

\makeatletter

\usepackage{hyperref}
\hypersetup{colorlinks = true,allcolors = blue}
\usepackage{times}

\makeatother

\usepackage{babel}
\begin{document}

\title{Quantum anomalous Hall effect in ferromagnetic metals}
\author{Yu-Hao Wan}
\affiliation{International Center for Quantum Materials and School of Physics,
Peking University, Beijing 100871, China}
\author{Peng-Yi Liu}
\affiliation{International Center for Quantum Materials and School of Physics,
Peking University, Beijing 100871, China}
\author{Qing-Feng Sun}
\email[Corresponding author: ]{sunqf@pku.edu.cn.}
\affiliation{International Center for Quantum Materials and School of Physics, Peking University, Beijing 100871, China}
\affiliation{Hefei National Laboratory, Hefei 230088, China}

\begin{abstract}
The quantum anomalous Hall (QAH) effect holds fundamental importance
in topological physics and technological promise for electronics.
It is generally believed that the QAH effect can only be realized in insulators.
In this work, we theoretically demonstrate that the QAH effect can also be realized in metallic systems, representing a phase distinct from the conventional QAH phase in insulators. This phase is characterized by the coexistence
of chiral edge channels and isotropic bulk conduction channels without
a bulk energy gap.
Notably, in a six-terminal Hall bar, our calculations show that, the quantized Hall conductivity and nonzero longitudinal conductivity can emerge due to dephasing, despite the Hall resistivity itself never becoming quantized.
Furthermore, the quantized Hall conductivity exhibits remarkable robustness against disorder.
Our findings not only extend the range of materials capable of hosting the QAH effect
from insulators to metals, but also provide insights that may pave
the way for the experimental realization of the QAH effect at elevated
temperatures.
\end{abstract}

\maketitle

$\textit{Introduction}$.\textemdash
The QAH effect, as a representative topological phase of matter, has attracted widespread attention in recent years due to its remarkable property: quantized Hall conductivity in the absence of an external magnetic field \citep{chang_colloquium_2023,chang_experimental_2013,deng_quantum_2020,chen_tunable_2020,liu_quantum_2008,haldane_model_1988}.
Unlike the conventional Hall effect, where quantization relies on
strong magnetic fields, the QAH effect arises from intrinsic spin-orbit
coupling and magnetization, leading to robust quantized Hall conductance
that is insensitive to disorder \citep{jiang_numerical_2009}.
Traditionally,
the QAH effect has been realized and studied mostly in ferromagnetic insulators,
where a well-defined global bulk energy gap supports dissipationless chiral edge states \citep{chang_high-precision_2015,checkelsky_trajectory_2014,chang_experimental_2013,kou_scale-invariant_2014,okazaki_quantum_2022,deng_quantum_2020,chen_topological_2019,chen_intrinsic_2019,liu_robust_2020,liang_chern_2025}.
The number of these edge states is determined by the Chern number $\mathbb{C}$, a topological invariant of the system \citep{bernevig_topological_2013}.
Such ferromagnetic insulators are very rare in nature \citep{checkelsky_trajectory_2014,chang_colloquium_2023}, and the Curie temperature is usually extremely low \cite{lee_strong_2010}, which severely limits the experimental and applied research of QAH effect.

% Recent work demonstrated quantized Hall conductance in a metallic regime by overlapping the surface states of two semi-magnetic topological insulators \citep{bai_metallic_2023}. Although the bulk density of states is metallic, the quantization originates from the underlying insulating topology of the three-dimensional topological insulators. This leaves open the question whether a truly intrinsic QAH phase can arise in purely two-dimensional ferromagnetic metals without a bulk gap.

Recent work demonstrated quantized Hall conductance in a topological insulator film with a magnetic sandwich heterostructure, where gapless Dirac surface states persist \citep{bai_metallic_2023}. While the quantization stems from the parity anomaly of these surface states, their existence ultimately depends on the insulating topology of the three-dimensional topological insulators\citep{mogi_experimental_2022,fu2022quantum,zou2022half,zou2023half,ning2023robustness,wang2024signature,wan_quarter-quantized_2024,fu2024half}. This raises the question of whether a truly intrinsic QAH phase can exist in purely two-dimensional ferromagnetic metals without a bulk gap.

Notably, ferromagnetic metals are widely present \cite{nagaosa_anomalous_2010}, and the Curie temperature often reaches several hundred Kelvin.
This discrepancy raises an intriguing and fundamental question:
Is it possible to realize the QAH effect in ferromagnetic metals, which
typically lack a global energy gap?
Exploring this question not only broadens the material scope for QAH physics but may also pave the way for novel quantum devices based on topological metallic phases.
This forms the central motivation of our study.

\begin{figure}
\begin{centering}
\includegraphics[scale=0.4]{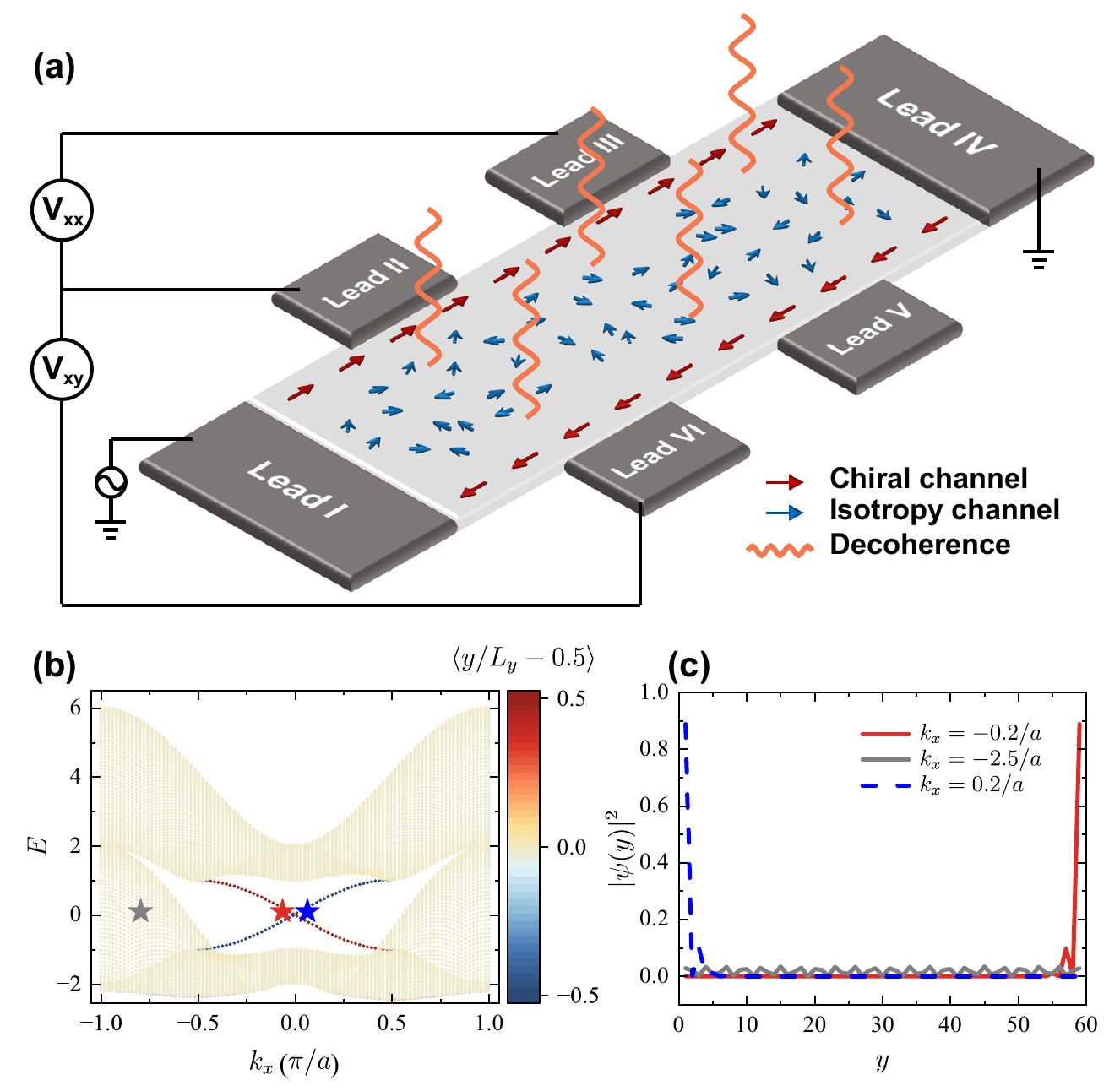}
\par\end{centering}
\caption{\label{fig:1}
(a) Schematic of the metallic QAH system under dephasing, showing coexisting isotropic (blue arrows) and chiral (red arrows) conduction channels.
%It is also a schematic diagram of a six-terminal Hall bar measuring the QAH effect, where
The dark gray regions represent the leads I-VI for measuring the QAH effect.
The size of the system is $L_{x}=240$ and $L_{y}=60$. The separation between leads
II (V) and III (VI) is $l=40$ (A schematic of the device geometry is shown in the Supplementary Materials\cite{seesupplemental}).
(b) Band structure of the nanoribbon system, with colors indicating the average displacement of each Bloch state relative to the center, $\langle y/L_{y}-1/2\rangle$.
(c) Spatial distribution of the states marked by differently colored
spheres in (b).}
\end{figure}

In this Letter, we theoretically propose and characterize a novel
topological phase that generalizes the QAH effect from insulating
to metallic systems. We construct a minimal model exhibiting both
chiral edge states and a conducting metallic bulk, and quantitatively
analyze its transport properties using a six-terminal Hall bar
setup [shown in Fig. \ref{fig:1}(a)] with dephasing.
Remarkably, we find that the Hall conductivity
$\sigma_{xy}$ remains quantized, forming a robust plateau even in
the presence of strong dephasing and disorder, while the longitudinal
conductivity $\sigma_{xx}$ remains finite. This robust quantization
%,which persists even under disorder,
directly reflects the topological protection originating from
band inversion and the nontrivial band topology of the system.
Our findings offer a theoretical foundation
for realizing QAH effects in metallic phases and open new avenues
for exploring topological phenomena in ferromagnetic metals.

\textit{}

\emph{Model Hamiltonian}.~\textemdash To capture the essential physics of the QAH effect in ferromagnetic metals, we consider a minimal three-band tight-binding model with broken time-reversal symmetry, as would arise in a ferromagnetic system:
\[
H(\mathbf{k}) = \begin{pmatrix}
\epsilon_{k_{1}} & t & t_{soc} \\
t^{*} & \epsilon_{k_{2}} & t \\
t_{soc}^{*} & t^{*} & \epsilon_{k_{3}}
\end{pmatrix},
\]

written in a minimal three orbital basis ($|1\rangle$,$|2\rangle$,$|3\rangle$), where $|1\rangle$ and $|3\rangle$ are two exchange-split orbitals in a ferromagnetic metal with a band inversion between them, while $|2\rangle$ is a intermediate conduction band that can be viewed as an effective spin-polarized band, as commonly found in ferromagnetic metals. To further illustrate this physical origin, a four-band extension where the metallic channel explicitly arises from a spin-split pair is presented in the Supplementary Material\cite{seesupplemental}. The elements are defined as
\(\epsilon_{k_1} = -\epsilon_{k_3}= m - 2B[2 - (\cos k_x + \cos k_y)]\),
%\(\epsilon_{k_3} = -[m - 2B[2 - (\cos k_x + \cos k_y)]]\),
\(\epsilon_{k_2} = 2t_0(\cos k_x + \cos k_y) + \varepsilon_0\), and \(t_{soc} = A(\sin k_x - i \sin k_y)\).
Here, the effective magnetization is incorporated through the mass term $m$ and the associated band inversion, which mimics the exchange splitting and spin polarization characteristic of ferromagnetic metals with strong spin-orbit coupling. The parameter \(t\) denotes the hybridization strength between different orbitals, and \(t_{soc}\) represents the spin-orbit coupling.
In our calculations, we set \(A = B = 1\), \(m = 2\), \(t = 0.2\), \(\varepsilon_{0} = -2.8\), \(t_0 = 1.2\) and use these values unless otherwise specified. This parameter choice ensures that the system resides in the metallic QAH phase, where robust chiral edge states coexist with finite bulk conductivity. Further details of the parameter constraints required for realizing this phase are provided in Supplementary Material \cite{seesupplemental}.

To see this, we construct the corresponding tight-binding Hamiltonian in a ribbon geometry with width $L_{y}=60a$, where \ensuremath{a} is the lattice constant (set $a=1$ throughout this work), while maintaining translational symmetry along the $x$-direction.
Fig. \ref{fig:1}(b) shows the band structure of this nanoribbon.
The absence of a full energy gap indicates that the system is metallic. However, in contrast to an ordinary metal, the band structure exhibits a clear band inversion, which gives rise to the presence of chiral edge modes \cite{qi_topological_2006,bernevig_topological_2013}.
To visualize the spatial character of each Bloch state, we compute
the expectation value $\langle y/L_{y}-1/2\rangle$ and color-code
the states according to their average displacement from the ribbon
center.
States localized near the left (red) and right (blue) edges possess opposite group velocities, confirming the presence of chiral
edge channels.
This is further illustrated by plotting representative real-space wave functions at three selected momenta: the red and blue curves are sharply localized at opposite edges (signatures of chiral edge modes), while the grey curve is extended throughout the ribbon, indicative of a bulk metallic state.
This clearly demonstrates the simultaneous presence of %topologically protected
chiral edge states and a metallic bulk in our model.

The appearance of chiral edge modes suggests the existence of quantized Hall conductivity.
However, in the experiment, the Hall conductivity is not directly measurable but is converted from resistance \cite{mogi_experimental_2022,zhou_transport_2022}.
Fig.\ref{fig:1}(a) shows a standard Hall bar geometry commonly used in experiments to measure the QAH effect. Lead I serves as the current source, injecting current \(I\) into the device, while lead IV acts as the drain and is grounded. Leads II, III, V, and VI function as voltage probes, measuring the local potentials without drawing net current.
The longitudinal voltage $V_{xx}$ (Hall voltage $V_{xy}$) is measured by leads II and III (II and VI).
Corresponding resistivities are given by, $\rho_{xx}=R_{xx}/(L_{x}/L_{y})=V_{xx}L_y / I L_x$ and $\rho_{xy}=R_{xy}=V_{xy}/I$ \cite{mogi_experimental_2022,zhou_transport_2022}.
Finally, the longitudinal and Hall conductivities are obtained by inverting the resistivity tensor:
\begin{equation}
\sigma_{xx}=\frac{\rho_{xx}}{\rho_{xx}^2+\rho_{xy}^2},\;\sigma_{xy}=\frac{\rho_{xy}}{\rho_{xx}^2+\rho_{xy}^2}\label{eq:conductivity}.
\end{equation}

Similar to the experiment, we also simulated the transport of the Hall bar in Fig. \ref{fig:1}(a) and converted the resistances into conductivities.
Besides six real metal leads, we also introduce B\"{u}ttiker virtual leads to incorporate dephasing effects at every $2\times2$ lattice site \citep{buttiker_four-terminal_1986,buttiker_symmetry_1988,buttiker_absence_1988,wan_quarter-quantized_2024,wan_altermagnetism-induced_2025,zhou_transport_2022,mclennan_voltage_1991,liu_dissipation_2024}.
The current of lead $n$ ($I_n$) can be calculated by Landauer-B\"{u}ttiker formula \citep{meir_landauer_1992,lambert_multi-probe_1993,liu_dissipation_2024,wan_mag}:
\begin{equation}
I_{n} = \frac{e^2}{h} \sum_{m \neq n} \left( T_{mn} V_n - T_{nm} V_m \right),
\label{eq:LB}
\end{equation}
where \( V_m \) is the voltage at lead \( m \), and \( T_{nm}(E) \) is the transmission coefficient from lead \( m \) to lead \( n \), evaluated as
$
T_{nm} = \operatorname{Tr}\left[\mathbf{\Gamma}_{n}(E_F) \mathbf{G}^{R}(E_F) \mathbf{\Gamma}_{m}(E_F) \mathbf{G}^{A}(E_F) \right]
$
\citep{meir_landauer_1992}. 
Here the Fermi energy $E_F$ is introduced as the chemical potential in the Landauer-B\"{u}ttiker formalism.
 \(\mathbf{\Gamma}_{n}(E)\) is the line-width function for lead \( n \). For leads I and IV, \(\mathbf{\Gamma}_{I/IV}(E) = i\left[\Sigma_{I/IV}^{R}(E) - \Sigma_{I/IV}^{A}(E)\right]\), where \(\Sigma_{I/IV}^{R/A}(E)\) are the corresponding retarded/advanced self-energies, which are calculated self-consistently using the iterative Green's function method. For the virtual leads, we set \(\mathbf{\Gamma}_{n} = \Gamma_d\) as a constant, where \(\Gamma_d\) characterizes the dephasing strength. The retarded Green's function is given by
$
\mathbf{G}^{R}(E) =\mathbf{G}^{A\dagger}(E) = \left[(E + i0^+)\mathbf{I} - \mathbf{H} - \sum_{n}\mathbf{\Sigma}_{n}^{R}(E)\right]^{-1},
$
where the total self-energy from all leads is included, and \(\mathbf{\Sigma}_{n}^{R}(E) = -\frac{i}{2}\mathbf{\Gamma}_{n}\) for the virtual leads.

\begin{figure}
\begin{centering}
\includegraphics[scale=0.15]{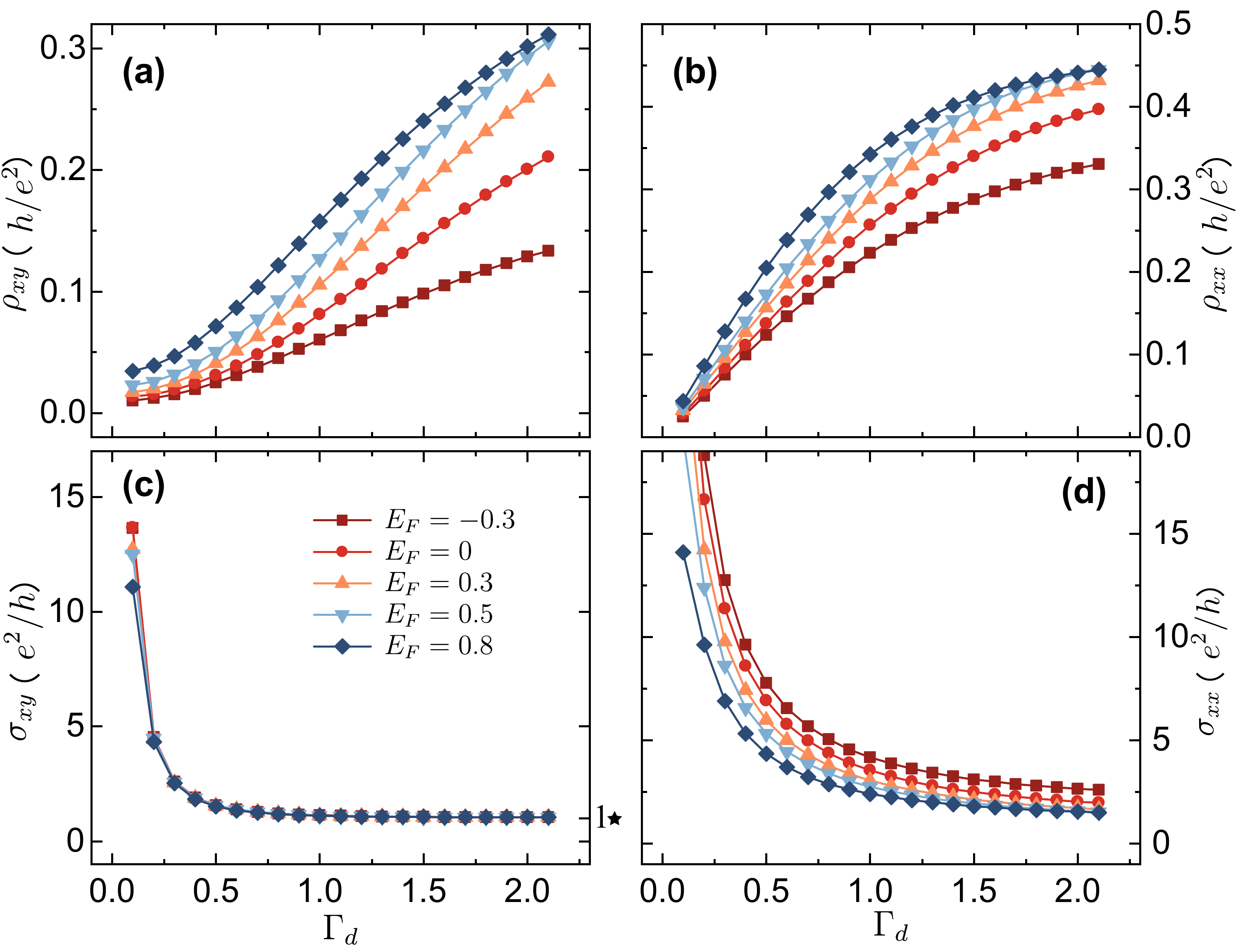}
\par\end{centering}
\caption{\label{fig:2}The Hall resistance $\rho_{xy}$ (a), longitudinal
resistance $\rho_{xx}$ (b), Hall conductance $\sigma_{xy}$ (c),
and longitudinal conductance $\sigma_{xx}$ (d) as functions of dephasing
strength $\Gamma_{d}$. The curves with different colors represent different Fermi energies $E_{F}$ as shown in (c).}
\end{figure}

According to Eq. (\ref{eq:LB}), we calculate the Hall and longitudinal resistances as in experimental measurements [$V_{\rm I}=1$ (is set to be the unit of voltages) and $V_{\rm IV}=0$], and further convert them into conductivities according to Eq. (\ref{eq:conductivity}), with results under different dephasing strengths $\Gamma_d$ shown in Fig. \ref{fig:2}.

\textit{Quantum Hall conductance under dephasing}\textemdash
In the QAH effect achieved in common insulators, the Hall resistance is quantized to $h/\mathbb{C}e^2$ and the longitudinal resistance disappears, along with the quantization of Hall conductivity to $\mathbb{C}e^2/h$ and disappearance of the longitudinal conductivity \cite{chang_colloquium_2023,chang_experimental_2013}.
However, in the metallic QAH effect, the Hall resistance has never been quantized under different conditions (Fermi energy $E_F$ and dephasing strength $\Gamma_d$) [Fig. \ref{fig:2}(a)], and the longitudinal resistance has never disappeared [Fig. \ref{fig:2}(b)] as a sign of the metal.
In the weak-dephasing limit ($\Gamma_d\ll1$), the Hall conductivity $\sigma_{xy}$ is nonzero but not quantized, which is due to the influence of ballistic transport of the metal as we will see below.
Surprisingly, as dephasing increases, although both $\rho_{xy}$ and $\rho_{xx}$ continue to vary with $\Gamma$, the Hall conductivity $\sigma_{xy}$ at various Fermi energies steadily approaches the quantized value $e^{2}/h$ [Fig. \ref{fig:2}(c)], precisely corresponding to the number of chiral edge states intersecting the Fermi level.
Meanwhile, the longitudinal conductivity $\sigma_{xx}$ remains finite [Fig. \ref{fig:2}(d)], indicating the remarkable coexistence of quantized Hall conductance with bulk metallic conduction. The decrease of $\sigma_{xx}$ with increasing $E_F$ originates from the reduction of propagating modes in the relevant energy window\cite{seesupplemental}.
Below, we will reveal what happens that leads to the quantization of $\sigma_{xy}$ as dephasing intensifies.

\begin{figure}
    \begin{centering}
    \includegraphics[scale=0.45]{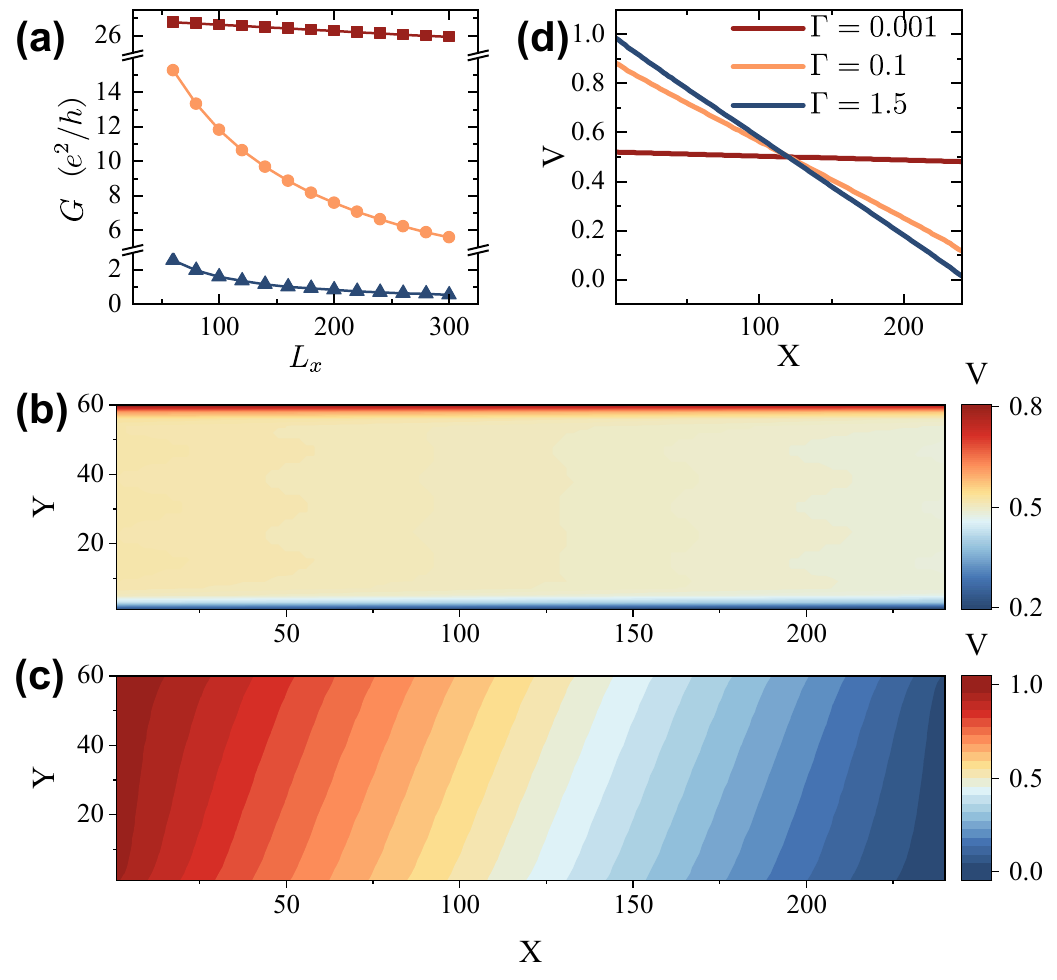}
    \par\end{centering}
    \caption{\label{fig:3}(a) Two-terminal conductance $G$ as a function of
    the central region length $L_{x}$ for various $\Gamma_d$.
    (b\textendash c) Spatial distribution of the potential for $\Gamma_d=0.001$ (b) and $\Gamma_d=1.5$ (c).
    (d) Potential distribution along the midline ($y=L_{y}/2$) of the
    central region for different $\Gamma_d$.
    $E_{F}=0$ in all calculations.
    }
\end{figure}

As the dephasing strength $\Gamma_d$ increases, the system undergoes
a crossover from a quantum (ballistic) regime to a classical (diffusive)
regime \cite{mclennan_voltage_1991}.
For small $\Gamma_d$, quantum effects dominate and the transport is nearly ballistic.
This is evident from the two-terminal conductance $G=I/(V_{\rm I}-V_{\rm IV})$ as a function of the central region length $L_{x}$ [see Fig. \ref{fig:3}(a)].
When $\Gamma_{d}=0.001$ (red curve), $G$ decreases only slightly as $L_{x}$ increases, a hallmark of ballistic transport and thus quantum coherence.
With increasing dephasing, $G$ decreases inversely with $L_{x}$, signaling the emergence of diffusive metallic bulk transport. %and the suppression of quantum effects.
This transition is further illustrated by the spatial potential distributions shown in Figs. \ref{fig:3} (b,c) for $\Gamma_d=0.001$ and $\Gamma_d=1.5$, respectively.
For weak dephasing ($\Gamma_d=0.001$), the potential across the central region remains nearly uniform (close to 1/2), which is characteristic of quantum ballistic transport.
At the same time, the upper (lower) edge of the sample has higher (lower) potential than the center, indicating that chiral edge states still exist when the Hall conductivity is not quantized.
In contrast, for strong dephasing ($\Gamma_d=1.5$), the potential exhibits a pronounced gradient, dropping smoothly from the biased left lead ($V_{\rm I}=$ 1) to the grounded right lead ($V_{\rm IV}=0$), reflecting the dominance of classical diffusive transport.
To further highlight the quantum-to-classical crossover induced by dephasing, Fig. \ref{fig:3}(d) presents the potential profile
along the midline ($y=L_{y}/2$) of the central region for different
values of $\Gamma_d$.
At small $\Gamma_d$, the potential is nearly flat, whereas for large $\Gamma_d$, it exhibits a continuous linear drop from one end to the other, providing clear evidence for the transition from quantum to classical metallic transport with increasing dephasing \cite{mclennan_voltage_1991}.

The persistence of quantized Hall conductivity $\sigma_{xy}=e^{2}/h$
under strong dephasing can be understood by modeling the system as
two parallel conduction channels \cite{mogi_experimental_2022}: (1) chiral edge modes,
which contribute a quantized Hall conductivity ($\sigma_{xy}^{C}=e^{2}/h$)
with vanishing longitudinal conductivity ($\sigma_{xx}^{C}=0$),
and (2) an isotropic metallic bulk, which provides only longitudinal
conduction ($\ensuremath{\sigma_{xx}^{M}>0}$) and no Hall response ($\sigma_{xy}^{M}=0$), since the metallic band is topologically trivial with negligible Berry curvature\cite{seesupplemental}.
In the presence of a longitudinal electric field $E_{x}$, the total
current density is $j_{x}=I/L_y=j_{x}^{M}+j_{x}^{C}=\sigma_{xx}^{M}E_{x}+\sigma_{xy}^{C}E_{y}$,
while the chiral channel contributes only via the transverse field
$E_{y}$.

Crucially, under strong dephasing, the isotropic metallic bulk enforces
local equilibrium, giving rise to a well-defined electric field throughout
the sample\textemdash a situation that does not arise in the quantum
(ballistic) regime, where the central region is field-free. The transverse
current of the sample is zero due to the open boundary \cite{mogi_experimental_2022}, resulting $0=j_{y}^{M}+j_{y}^{C}=\sigma_{xx}^{M}E_{y}+\sigma_{yx}^{C}E_{x}$.
Using $\sigma_{yx}^{C}=-\sigma_{xy}^{C}$, this gives $E_{y}=(\sigma_{xy}^{C}/\sigma_{xx}^{M})E_{x}$.
The measured Hall resistivity and conductivity are then, respectively,
$\rho_{xy}=L_yE_{y}/I$ and $\rho_{xx}=L_y R_{xx}/L_x = L_yE_{x}/I$.
Substituting the above relations into Eq. (\ref{eq:conductivity}), one finds that the total Hall conductivity remains quantized, $\sigma_{xy}=e^{2}/h$, regardless of the deviations in $\rho_{xy}$ and $\rho_{xx}$, which reproduces and explains the result in Fig. \ref{fig:2}.
In short, our analysis, which aligns well with our numerical results,
demonstrates that, despite the absence of a global bulk
gap and the presence of finite longitudinal conduction, quantized
Hall response can still be achieved, which offers a new paradigm
for QAH phenomena in metallic systems.

%In short, under strong dephasing, the edge and bulk channels decouple:
%the bulk mediates dissipation while the chiral edge modes preserve
%quantized Hall conduction. The dephasing-induced redistribution of
%the electric potential in the central region is thus essential for
%the emergence of quantized Hall conductivity in a metallic setting.
%Our analysis demonstrates that, despite the absence of a global bulk
%gap and the presence of finite longitudinal conduction, quantized
%Hall response can still be achieved, which offering a new paradigm
%for QAH phenomena in metallic systems.

It is also worth noting that the ``strong dephasing" we mentioned above does not mean that strict conditions are required to implement metallic QAH effect. 
Dephasing is ubiquitous in real devices\cite{youn2008nonequilibrium,chakravarty1986weak,stern1990phase,burkard1999coupled}.
In fact, to disrupt the ballistic transport of metals in the sample and achieve the potential distribution as shown in Fig. \ref{fig:3}(c), the coherence length needs to be less than the sample size $L_\phi < (L_x,L_y)$ \cite{mclennan_voltage_1991,addr1}.
In real experiments, the sample size is usually several micrometers to hundreds of micrometers.
Therefore, achieving appropriate coherence to simultaneously satisfy $L_\phi < (L_x,L_y)$ and maintaining the quantization of Hall conductivity is not difficult \cite{mclennan_voltage_1991}.
This has been commonly achieved in previous experiments on the quantum Hall effects \cite{huckestein_scaling_1995} and is very promising to be extended to metallic QAH system \cite{deng_probing_2022}. Consistently, our simulations reveal that $\sigma_{xy}$ quantization emerges in the same regime~\cite{seesupplemental}.

\begin{figure}
\begin{centering}
\includegraphics[width=\columnwidth]{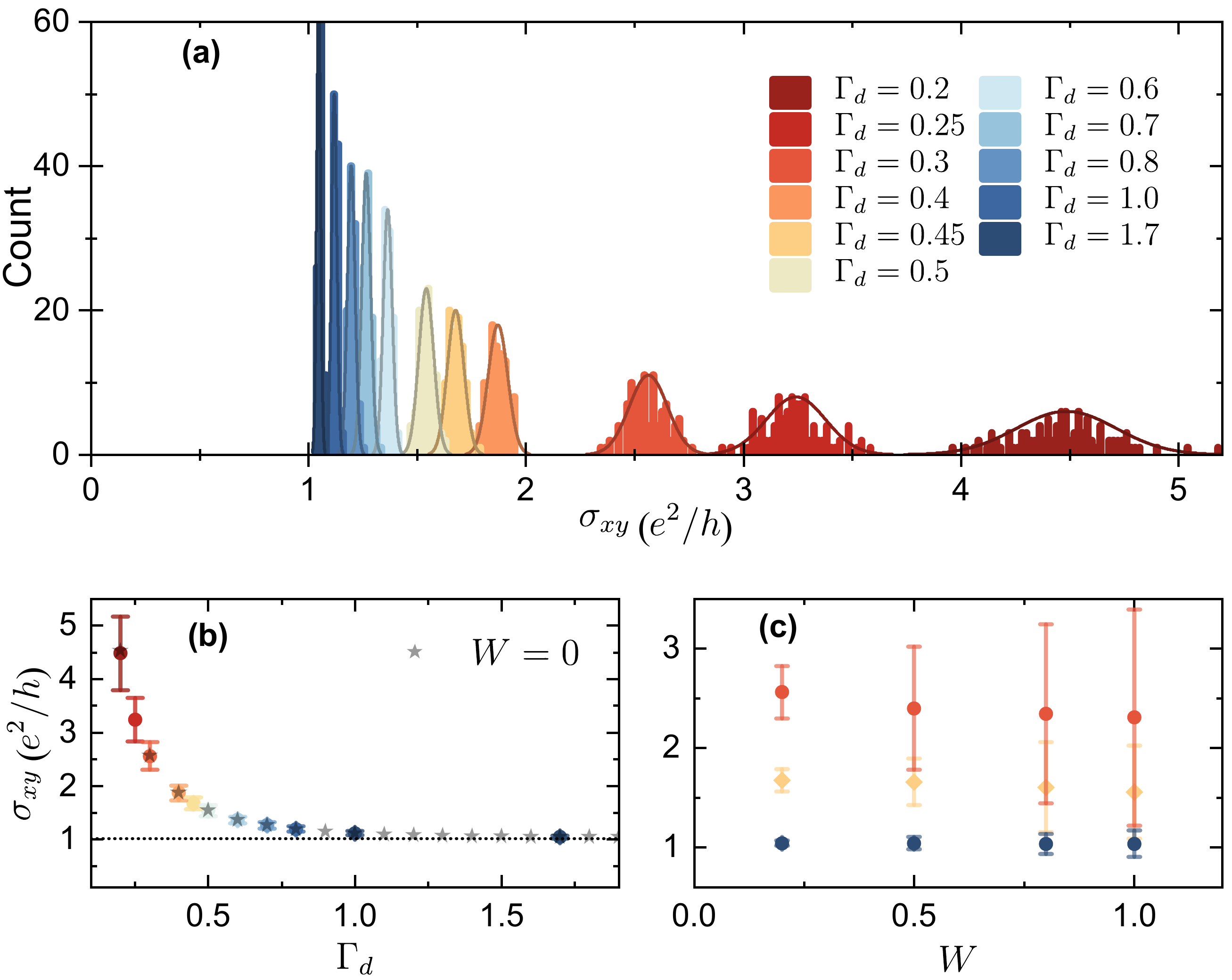}
\par\end{centering}
\caption{\label{fig:4}(a) Histogram of the frequency distribution of $\sigma_{xy}$
for various $\Gamma_d$, obtained from 100 random
configurations, with $W=0.2$. 
(b) The color stars represent the average $\sigma_{xy}$
over 100 random configuration sets and the gray stars correspond
to $\sigma_{xy}$ when $W=0$. (c) Dependence of $\sigma_{xy}$ on
increasing disorder for different dephasing strengths.  The red, yellow, and blue point correspond to $\Gamma_{d}=0.3, 0.45,$ and $1.7$, respectively.
The error bars in (b,c) indicate the range
$\left(\bar{\sigma}_{xy}-3\alpha,\,\bar{\sigma}_{xy}+3\alpha\right)$,
where $\bar{\sigma}_{xy}$ and $\alpha$ denote the mean and standard
deviation of the 100 data sets, respectively.
%The error bars again represent the range $\left(\bar{\sigma}_{xy}-3\alpha,\,\bar{\sigma}_{xy}+3\alpha\right)$.
In these calculations, we set $E_{F}=0$.}
\end{figure}

\emph{Stability under Disorder and Topological Protection}\textemdash{}.
Quantized phenomena are generally robust against disorder due to their
topological origin. Guided by this intuition, we systematically investigated
the disorder dependence of the Hall conductivity $\sigma_{xy}$ under
various dephasing strengths $\Gamma_d$.
The Anderson-type disorder is introduced via random on-site
energies uniformly distributed in $[-W/2,W/2]$ \cite{long_disorder-induced_2008,jiang_numerical_2009,liu_four-terminal_2024}, and the Hall response %at the Fermi energy $E_{F}=0$
is evaluated by Eq. (\ref{eq:conductivity},\ref{eq:LB}).
As shown in Fig. \ref{fig:4}(a), for weak dephasing, $\sigma_{xy}$ displays strong sample-to-sample fluctuations and deviates significantly from quantization.
However, as dephasing increases, the Gaussian-like distribution of $\sigma_{xy}$ gradually move and converges to the quantized value $e^{2}/h$.
Meanwhile, the width of the distribution narrows significantly.
This indicates that strong dephasing suppresses sample-to-sample fluctuations induced by disorder, reflecting the role of topological protection.

Figs. \ref{fig:4}(b,c) provide quantitative evidence for the robustness
of the Hall conductance in the presence of disorder and dephasing.
In Fig. \ref{fig:4}(b), we plot the average value of $\sigma_{xy}$
%(over 100 disorder realizations)
as a function of dephasing strength $\Gamma_d$ at disorder strength $W=0.2$.
%The error bars represent three times the standard deviation, thereby reflecting the spread of the Hall response due to disorder.
For weak dephasing $\Gamma_d$, not only does the average $\sigma_{xy}$ deviate from quantization, but the error bars are also large, indicating
strong sample-to-sample fluctuations.
As the $\Gamma_d$ increases, the average value rapidly approaches the quantized value $e^{2}/h$ and the error bars shrink, demonstrating that both the quantization and the stability of the Hall response are enhanced by dephasing, even for substantial disorder.
Fig. \ref{fig:4}(c) further explores the effect of increasing disorder strength $W$ for several fixed dephasing strengths ($\Gamma_d=0.3,0.5,1.7$).
For strong dephasing ($\Gamma_d=1.7$), the average $\sigma_{xy}$ remains
nearly quantized and the fluctuations remain small, even as the disorder is
increased up to $W=1$.
In contrast, for weak dephasing ($\Gamma_d=0.3$),
both the mean value and the stability of $\sigma_{xy}$ deteriorate
rapidly as disorder increases.
This result highlights that sufficiently strong dephasing is essential for maintaining quantized and robust Hall conductance in the metallic QAH system.

% Under strong dephasing, the remarkable robustness of the quantized
% Hall conductance in metallic QAH systems demonstrates that the quantized
% Hall plateau in our metallic QAH system is protected by the underlying
% band inversion and chiral edge states, i.e., it is a manifestation
% of intrinsic topological protection \cite{chang_colloquium_2023}.
% In conventional anomalous Hall systems, extrinsic contributions (such as skew scattering or side-jump mechanisms related to Fermi surface effects) lead to not quantized Hall conductivity \cite{nagaosa_anomalous_2010}.
% In our model, however, the observed quantization of $\sigma_{xy}$ is robustly maintained even in the presence of strong disorder and significant dephasing.
% This insensitivity to disorder and the persistence of quantization are key hallmarks of intrinsic, topologically protected transport, which is distinct from extrinsic contributions that are typically sensitive to details of impurity scattering and Fermi surface geometry.
% This intrinsic topological protection not only distinguishes our mechanism from extrinsic anomalous Hall effects, but also suggests that, in realistic experiments, the quantized plateau should be observable even in metallic QAH materials with unavoidable disorder and finite dephasing.
% The coexistence of chiral edge modes and metallic bulk states, protected by band inversion, constitutes a unique and experimentally accessible platform for exploring topological phenomena in ferromagnetic metals.
Under strong dephasing, the remarkable robustness of the quantized Hall conductance observed in our metallic QAH system demonstrates that the quantized Hall plateau originates from the underlying band inversion and the associated chiral edge states. This robustness against disorder and finite dephasing highlights the intrinsic topological nature of the observed quantization, distinguishing it fundamentally from conventional metallic Hall effects. The coexistence of robust chiral edge modes and metallic bulk states, stabilized by band inversion, offers a unique and experimentally accessible platform to explore topological phenomena in ferromagnetic metals.

\emph{Discussion and conclusion}.\textemdash{}
In summary, we have theoretically uncovered a new mechanism
for realizing the QAH effect in metallic systems.
By introducing dephasing into a minimal three-band tight-binding model that simultaneously hosts chiral edge states and a conducting bulk,
we demonstrated, using a six-terminal Landauer\textendash B\"{u}ttiker framework with virtual B\"{u}ttiker leads, that the Hall conductivity can become quantized to $e^{2}/h$, %within a finite energy window,
even in the absence of an insulating bulk gap and with a finite longitudinal conductance. Notably, this quantized Hall plateau is remarkably robust against strong Anderson disorder, highlighting the practical feasibility of the mechanism. Our work redefines the material landscape for QAH phenomena by demonstrating their feasibility in gapless ferromagnetic metals, thereby providing new directions for the search and design of QAH systems that are compatible with high-temperature operation.

% Our work broadens the material landscape for QAH physics by demonstrating the feasibility of quantized Hall effects in gapless metallic systems, and provides a theoretical basis for exploring QAH phenomena beyond insulating regimes.

% We note that a recent work \citep{bai_metallic_2023} proposed a so called ``metallic QAH phase" arising from a pair of massless Dirac surface states, each contributing half of the Hall conductance via quantum anomaly, which has been well understood \cite{haldane_model_1988,mogi_experimental_2022}.
% That scenario fundamentally relies on additive half-integer contributions from paired Dirac cones. %, without introducing qualitatively new physics.
% In contrast, our work reveals a distinct mechanism for metallic QAH
% behavior that does not require such paired Dirac cones, but instead
% arises from band inversion\textemdash a feature broadly present in
% various material systems. As a result, our theory is likely to be
% experimentally verifiable in a wide range of candidate materials.

$\textit{Acknowledgment}s$.\textemdash Y.-H. W. is grateful to Hu-Mian
Zhou for fruitful discussions.
This work was financially supported
by the National Key R and D Program of China (Grant No. 2024YFA1409002),
the National Natural Science Foundation of China (Grants No. 12374034 and Grants No. 124B2069),
and the Innovation Program for Quantum Science and Technology (Grant No. 2021ZD0302403).
We acknowledge the High-performance Computing Platform of Peking University
for providing computational resources.

\bibliography{ref}

\begin{thebibliography}{48}%
\makeatletter
\providecommand \@ifxundefined [1]{%
 \@ifx{#1\undefined}
}%
\providecommand \@ifnum [1]{%
 \ifnum #1\expandafter \@firstoftwo
 \else \expandafter \@secondoftwo
 \fi
}%
\providecommand \@ifx [1]{%
 \ifx #1\expandafter \@firstoftwo
 \else \expandafter \@secondoftwo
 \fi
}%
\providecommand \natexlab [1]{#1}%
\providecommand \enquote  [1]{``#1''}%
\providecommand \bibnamefont  [1]{#1}%
\providecommand \bibfnamefont [1]{#1}%
\providecommand \citenamefont [1]{#1}%
\providecommand \href@noop [0]{\@secondoftwo}%
\providecommand \href [0]{\begingroup \@sanitize@url \@href}%
\providecommand \@href[1]{\@@startlink{#1}\@@href}%
\providecommand \@@href[1]{\endgroup#1\@@endlink}%
\providecommand \@sanitize@url [0]{\catcode `\\12\catcode `\$12\catcode `\&12\catcode `\#12\catcode `\^12\catcode `\_12\catcode `\%12\relax}%
\providecommand \@@startlink[1]{}%
\providecommand \@@endlink[0]{}%
\providecommand \url  [0]{\begingroup\@sanitize@url \@url }%
\providecommand \@url [1]{\endgroup\@href {#1}{\urlprefix }}%
\providecommand \urlprefix  [0]{URL }%
\providecommand \Eprint [0]{\href }%
\providecommand \doibase [0]{https://doi.org/}%
\providecommand \selectlanguage [0]{\@gobble}%
\providecommand \bibinfo  [0]{\@secondoftwo}%
\providecommand \bibfield  [0]{\@secondoftwo}%
\providecommand \translation [1]{[#1]}%
\providecommand \BibitemOpen [0]{}%
\providecommand \bibitemStop [0]{}%
\providecommand \bibitemNoStop [0]{.\EOS\space}%
\providecommand \EOS [0]{\spacefactor3000\relax}%
\providecommand \BibitemShut  [1]{\csname bibitem#1\endcsname}%
\let\auto@bib@innerbib\@empty
%</preamble>
\bibitem [{\citenamefont {Chang}\ \emph {et~al.}(2023)\citenamefont {Chang}, \citenamefont {Liu},\ and\ \citenamefont {MacDonald}}]{chang_colloquium_2023}%
  \BibitemOpen
  \bibfield  {author} {\bibinfo {author} {\bibfnamefont {C.-Z.}\ \bibnamefont {Chang}}, \bibinfo {author} {\bibfnamefont {C.-X.}\ \bibnamefont {Liu}},\ and\ \bibinfo {author} {\bibfnamefont {A.~H.}\ \bibnamefont {MacDonald}},\ }\bibfield  {title} {\bibinfo {title} {Colloquium: {Quantum} anomalous {Hall} effect},\ }\href {https://doi.org/10.1103/RevModPhys.95.011002} {\bibfield  {journal} {\bibinfo  {journal} {Rev. Mod. Phys.}\ }\textbf {\bibinfo {volume} {95}},\ \bibinfo {pages} {011002} (\bibinfo {year} {2023})}\BibitemShut {NoStop}%
\bibitem [{\citenamefont {Chang}\ \emph {et~al.}(2013)\citenamefont {Chang}, \citenamefont {Zhang}, \citenamefont {Feng}, \citenamefont {Shen}, \citenamefont {Zhang}, \citenamefont {Guo}, \citenamefont {Li}, \citenamefont {Ou}, \citenamefont {Wei}, \citenamefont {Wang}, \citenamefont {Ji}, \citenamefont {Feng}, \citenamefont {Ji}, \citenamefont {Chen}, \citenamefont {Jia}, \citenamefont {Dai}, \citenamefont {Fang}, \citenamefont {Zhang}, \citenamefont {He}, \citenamefont {Wang}, \citenamefont {Lu}, \citenamefont {Ma},\ and\ \citenamefont {Xue}}]{chang_experimental_2013}%
  \BibitemOpen
  \bibfield  {author} {\bibinfo {author} {\bibfnamefont {C.-Z.}\ \bibnamefont {Chang}}, \bibinfo {author} {\bibfnamefont {J.}~\bibnamefont {Zhang}}, \bibinfo {author} {\bibfnamefont {X.}~\bibnamefont {Feng}}, \bibinfo {author} {\bibfnamefont {J.}~\bibnamefont {Shen}}, \bibinfo {author} {\bibfnamefont {Z.}~\bibnamefont {Zhang}}, \bibinfo {author} {\bibfnamefont {M.}~\bibnamefont {Guo}}, \bibinfo {author} {\bibfnamefont {K.}~\bibnamefont {Li}}, \bibinfo {author} {\bibfnamefont {Y.}~\bibnamefont {Ou}}, \bibinfo {author} {\bibfnamefont {P.}~\bibnamefont {Wei}}, \bibinfo {author} {\bibfnamefont {L.-L.}\ \bibnamefont {Wang}}, \bibinfo {author} {\bibfnamefont {Z.-Q.}\ \bibnamefont {Ji}}, \bibinfo {author} {\bibfnamefont {Y.}~\bibnamefont {Feng}}, \bibinfo {author} {\bibfnamefont {S.}~\bibnamefont {Ji}}, \bibinfo {author} {\bibfnamefont {X.}~\bibnamefont {Chen}}, \bibinfo {author} {\bibfnamefont {J.}~\bibnamefont {Jia}}, \bibinfo {author} {\bibfnamefont {X.}~\bibnamefont {Dai}}, \bibinfo {author} {\bibfnamefont {Z.}~\bibnamefont {Fang}}, \bibinfo {author} {\bibfnamefont {S.-C.}\ \bibnamefont {Zhang}}, \bibinfo {author} {\bibfnamefont {K.}~\bibnamefont {He}}, \bibinfo {author} {\bibfnamefont {Y.}~\bibnamefont {Wang}}, \bibinfo {author} {\bibfnamefont {L.}~\bibnamefont {Lu}}, \bibinfo {author} {\bibfnamefont {X.-C.}\ \bibnamefont {Ma}},\ and\ \bibinfo {author} {\bibfnamefont {Q.-K.}\ \bibnamefont {Xue}},\ }\bibfield  {title} {\bibinfo {title} {Experimental {Observation} of the {Quantum} {Anomalous} {Hall} {Effect} in a {Magnetic} {Topological} {Insulator}},\ }\href {https://doi.org/10.1126/science.1234414} {\bibfield  {journal} {\bibinfo  {journal} {Science}\ }\textbf {\bibinfo {volume} {340}},\ \bibinfo {pages} {167} (\bibinfo {year} {2013})}\BibitemShut {NoStop}%
\bibitem [{\citenamefont {Deng}\ \emph {et~al.}(2020)\citenamefont {Deng}, \citenamefont {Yu}, \citenamefont {Shi}, \citenamefont {Guo}, \citenamefont {Xu}, \citenamefont {Wang}, \citenamefont {Chen},\ and\ \citenamefont {Zhang}}]{deng_quantum_2020}%
  \BibitemOpen
  \bibfield  {author} {\bibinfo {author} {\bibfnamefont {Y.}~\bibnamefont {Deng}}, \bibinfo {author} {\bibfnamefont {Y.}~\bibnamefont {Yu}}, \bibinfo {author} {\bibfnamefont {M.~Z.}\ \bibnamefont {Shi}}, \bibinfo {author} {\bibfnamefont {Z.}~\bibnamefont {Guo}}, \bibinfo {author} {\bibfnamefont {Z.}~\bibnamefont {Xu}}, \bibinfo {author} {\bibfnamefont {J.}~\bibnamefont {Wang}}, \bibinfo {author} {\bibfnamefont {X.~H.}\ \bibnamefont {Chen}},\ and\ \bibinfo {author} {\bibfnamefont {Y.}~\bibnamefont {Zhang}},\ }\bibfield  {title} {\bibinfo {title} {Quantum {Anomalous} {Hall} {Effect} in {Intrinsic} {Magnetic} {Topological} {Insulator} {MnBi2Te4}},\ }\href {https://doi.org/10.1126/science.aax8156} {\bibfield  {journal} {\bibinfo  {journal} {Science}\ }\textbf {\bibinfo {volume} {367}},\ \bibinfo {pages} {895} (\bibinfo {year} {2020})}\BibitemShut {NoStop}%
\bibitem [{\citenamefont {Chen}\ \emph {et~al.}(2020)\citenamefont {Chen}, \citenamefont {Sharpe}, \citenamefont {Fox}, \citenamefont {Zhang}, \citenamefont {Wang}, \citenamefont {Jiang}, \citenamefont {Lyu}, \citenamefont {Li}, \citenamefont {Watanabe}, \citenamefont {Taniguchi}, \citenamefont {Shi}, \citenamefont {Senthil}, \citenamefont {Goldhaber-Gordon}, \citenamefont {Zhang},\ and\ \citenamefont {Wang}}]{chen_tunable_2020}%
  \BibitemOpen
  \bibfield  {author} {\bibinfo {author} {\bibfnamefont {G.}~\bibnamefont {Chen}}, \bibinfo {author} {\bibfnamefont {A.~L.}\ \bibnamefont {Sharpe}}, \bibinfo {author} {\bibfnamefont {E.~J.}\ \bibnamefont {Fox}}, \bibinfo {author} {\bibfnamefont {Y.-H.}\ \bibnamefont {Zhang}}, \bibinfo {author} {\bibfnamefont {S.}~\bibnamefont {Wang}}, \bibinfo {author} {\bibfnamefont {L.}~\bibnamefont {Jiang}}, \bibinfo {author} {\bibfnamefont {B.}~\bibnamefont {Lyu}}, \bibinfo {author} {\bibfnamefont {H.}~\bibnamefont {Li}}, \bibinfo {author} {\bibfnamefont {K.}~\bibnamefont {Watanabe}}, \bibinfo {author} {\bibfnamefont {T.}~\bibnamefont {Taniguchi}}, \bibinfo {author} {\bibfnamefont {Z.}~\bibnamefont {Shi}}, \bibinfo {author} {\bibfnamefont {T.}~\bibnamefont {Senthil}}, \bibinfo {author} {\bibfnamefont {D.}~\bibnamefont {Goldhaber-Gordon}}, \bibinfo {author} {\bibfnamefont {Y.}~\bibnamefont {Zhang}},\ and\ \bibinfo {author} {\bibfnamefont {F.}~\bibnamefont {Wang}},\ }\bibfield  {title} {\bibinfo {title} {Tunable {Correlated} {Chern} {Insulator} and {Ferromagnetism} in a {Moir\'{e} {Superlattice}}},\ }\href {https://doi.org/10.1038/s41586-020-2049-7} {\bibfield  {journal} {\bibinfo  {journal} {Nature}\ }\textbf {\bibinfo {volume} {579}},\ \bibinfo {pages} {56} (\bibinfo {year} {2020})}\BibitemShut {NoStop}%
\bibitem [{\citenamefont {Liu}\ \emph {et~al.}(2008)\citenamefont {Liu}, \citenamefont {Qi}, \citenamefont {Dai}, \citenamefont {Fang},\ and\ \citenamefont {Zhang}}]{liu_quantum_2008}%
  \BibitemOpen
  \bibfield  {author} {\bibinfo {author} {\bibfnamefont {C.-X.}\ \bibnamefont {Liu}}, \bibinfo {author} {\bibfnamefont {X.-L.}\ \bibnamefont {Qi}}, \bibinfo {author} {\bibfnamefont {X.}~\bibnamefont {Dai}}, \bibinfo {author} {\bibfnamefont {Z.}~\bibnamefont {Fang}},\ and\ \bibinfo {author} {\bibfnamefont {S.-C.}\ \bibnamefont {Zhang}},\ }\bibfield  {title} {\bibinfo {title} {Quantum anomalous hall effect in hg1-ymnyte quantum wells},\ }\href {https://doi.org/10.1103/PhysRevLett.101.146802} {\bibfield  {journal} {\bibinfo  {journal} {Phys. Rev. Lett.}\ }\textbf {\bibinfo {volume} {101}},\ \bibinfo {pages} {146802} (\bibinfo {year} {2008})}\BibitemShut {NoStop}%
\bibitem [{\citenamefont {Haldane}(1988)}]{haldane_model_1988}%
  \BibitemOpen
  \bibfield  {author} {\bibinfo {author} {\bibfnamefont {F.~D.~M.}\ \bibnamefont {Haldane}},\ }\bibfield  {title} {\bibinfo {title} {Model for a {Quantum} {Hall} {Effect} without {Landau} {Levels}: {Condensed}-{Matter} {Realization} of the "{Parity} {Anomaly}"},\ }\href {https://doi.org/10.1103/PhysRevLett.61.2015} {\bibfield  {journal} {\bibinfo  {journal} {Phys. Rev. Lett.}\ }\textbf {\bibinfo {volume} {61}},\ \bibinfo {pages} {2015} (\bibinfo {year} {1988})}\BibitemShut {NoStop}%
\bibitem [{\citenamefont {Jiang}\ \emph {et~al.}(2009)\citenamefont {Jiang}, \citenamefont {Wang}, \citenamefont {Sun},\ and\ \citenamefont {Xie}}]{jiang_numerical_2009}%
  \BibitemOpen
  \bibfield  {author} {\bibinfo {author} {\bibfnamefont {H.}~\bibnamefont {Jiang}}, \bibinfo {author} {\bibfnamefont {L.}~\bibnamefont {Wang}}, \bibinfo {author} {\bibfnamefont {Q.-f.}\ \bibnamefont {Sun}},\ and\ \bibinfo {author} {\bibfnamefont {X.~C.}\ \bibnamefont {Xie}},\ }\bibfield  {title} {\bibinfo {title} {Numerical {Study} of the {Topological} {Anderson} {Insulator} in {HgTe}/{CdTe} {Quantum} {Wells}},\ }\href {https://doi.org/10.1103/PhysRevB.80.165316} {\bibfield  {journal} {\bibinfo  {journal} {Phys. Rev. B}\ }\textbf {\bibinfo {volume} {80}},\ \bibinfo {pages} {165316} (\bibinfo {year} {2009})}\BibitemShut {NoStop}%
\bibitem [{\citenamefont {Chang}\ \emph {et~al.}(2015)\citenamefont {Chang}, \citenamefont {Zhao}, \citenamefont {Kim}, \citenamefont {Zhang}, \citenamefont {Assaf}, \citenamefont {Heiman}, \citenamefont {Zhang}, \citenamefont {Liu}, \citenamefont {Chan},\ and\ \citenamefont {Moodera}}]{chang_high-precision_2015}%
  \BibitemOpen
  \bibfield  {author} {\bibinfo {author} {\bibfnamefont {C.-Z.}\ \bibnamefont {Chang}}, \bibinfo {author} {\bibfnamefont {W.}~\bibnamefont {Zhao}}, \bibinfo {author} {\bibfnamefont {D.~Y.}\ \bibnamefont {Kim}}, \bibinfo {author} {\bibfnamefont {H.}~\bibnamefont {Zhang}}, \bibinfo {author} {\bibfnamefont {B.~A.}\ \bibnamefont {Assaf}}, \bibinfo {author} {\bibfnamefont {D.}~\bibnamefont {Heiman}}, \bibinfo {author} {\bibfnamefont {S.-C.}\ \bibnamefont {Zhang}}, \bibinfo {author} {\bibfnamefont {C.}~\bibnamefont {Liu}}, \bibinfo {author} {\bibfnamefont {M.~H.~W.}\ \bibnamefont {Chan}},\ and\ \bibinfo {author} {\bibfnamefont {J.~S.}\ \bibnamefont {Moodera}},\ }\bibfield  {title} {\bibinfo {title} {High-{Precision} {Realization} of {Robust} {Quantum} {Anomalous} {Hall} {State} in a {Hard} {Ferromagnetic} {Topological} {Insulator}},\ }\href {https://doi.org/10.1038/nmat4204} {\bibfield  {journal} {\bibinfo  {journal} {Nat. Mater.}\ }\textbf {\bibinfo {volume} {14}},\ \bibinfo {pages} {473} (\bibinfo {year} {2015})}\BibitemShut {NoStop}%
\bibitem [{\citenamefont {Checkelsky}\ \emph {et~al.}(2014)\citenamefont {Checkelsky}, \citenamefont {Yoshimi}, \citenamefont {Tsukazaki}, \citenamefont {Takahashi}, \citenamefont {Kozuka}, \citenamefont {Falson}, \citenamefont {Kawasaki},\ and\ \citenamefont {Tokura}}]{checkelsky_trajectory_2014}%
  \BibitemOpen
  \bibfield  {author} {\bibinfo {author} {\bibfnamefont {J.~G.}\ \bibnamefont {Checkelsky}}, \bibinfo {author} {\bibfnamefont {R.}~\bibnamefont {Yoshimi}}, \bibinfo {author} {\bibfnamefont {A.}~\bibnamefont {Tsukazaki}}, \bibinfo {author} {\bibfnamefont {K.~S.}\ \bibnamefont {Takahashi}}, \bibinfo {author} {\bibfnamefont {Y.}~\bibnamefont {Kozuka}}, \bibinfo {author} {\bibfnamefont {J.}~\bibnamefont {Falson}}, \bibinfo {author} {\bibfnamefont {M.}~\bibnamefont {Kawasaki}},\ and\ \bibinfo {author} {\bibfnamefont {Y.}~\bibnamefont {Tokura}},\ }\bibfield  {title} {\bibinfo {title} {Trajectory of the {Anomalous} {Hall} {Effect} towards the {Quantized} {State} in a {Ferromagnetic} {Topological} {Insulator}},\ }\href {https://doi.org/10.1038/nphys3053} {\bibfield  {journal} {\bibinfo  {journal} {Nat. Phys.}\ }\textbf {\bibinfo {volume} {10}},\ \bibinfo {pages} {731} (\bibinfo {year} {2014})}\BibitemShut {NoStop}%
\bibitem [{\citenamefont {Kou}\ \emph {et~al.}(2014)\citenamefont {Kou}, \citenamefont {Guo}, \citenamefont {Fan}, \citenamefont {Pan}, \citenamefont {Lang}, \citenamefont {Jiang}, \citenamefont {Shao}, \citenamefont {Nie}, \citenamefont {Murata}, \citenamefont {Tang}, \citenamefont {Wang}, \citenamefont {He}, \citenamefont {Lee}, \citenamefont {Lee},\ and\ \citenamefont {Wang}}]{kou_scale-invariant_2014}%
  \BibitemOpen
  \bibfield  {author} {\bibinfo {author} {\bibfnamefont {X.}~\bibnamefont {Kou}}, \bibinfo {author} {\bibfnamefont {S.-T.}\ \bibnamefont {Guo}}, \bibinfo {author} {\bibfnamefont {Y.}~\bibnamefont {Fan}}, \bibinfo {author} {\bibfnamefont {L.}~\bibnamefont {Pan}}, \bibinfo {author} {\bibfnamefont {M.}~\bibnamefont {Lang}}, \bibinfo {author} {\bibfnamefont {Y.}~\bibnamefont {Jiang}}, \bibinfo {author} {\bibfnamefont {Q.}~\bibnamefont {Shao}}, \bibinfo {author} {\bibfnamefont {T.}~\bibnamefont {Nie}}, \bibinfo {author} {\bibfnamefont {K.}~\bibnamefont {Murata}}, \bibinfo {author} {\bibfnamefont {J.}~\bibnamefont {Tang}}, \bibinfo {author} {\bibfnamefont {Y.}~\bibnamefont {Wang}}, \bibinfo {author} {\bibfnamefont {L.}~\bibnamefont {He}}, \bibinfo {author} {\bibfnamefont {T.-K.}\ \bibnamefont {Lee}}, \bibinfo {author} {\bibfnamefont {W.-L.}\ \bibnamefont {Lee}},\ and\ \bibinfo {author} {\bibfnamefont {K.~L.}\ \bibnamefont {Wang}},\ }\bibfield  {title} {\bibinfo {title} {Scale-{Invariant} {Quantum} {Anomalous} {Hall} {Effect} in {Magnetic} {Topological} {Insulators} beyond the {Two}-{Dimensional} {Limit}},\ }\href {https://doi.org/10.1103/PhysRevLett.113.137201} {\bibfield  {journal} {\bibinfo  {journal} {Phys. Rev. Lett.}\ }\textbf {\bibinfo {volume} {113}},\ \bibinfo {pages} {137201} (\bibinfo {year} {2014})}\BibitemShut {NoStop}%
\bibitem [{\citenamefont {Okazaki}\ \emph {et~al.}(2022)\citenamefont {Okazaki}, \citenamefont {Oe}, \citenamefont {Kawamura}, \citenamefont {Yoshimi}, \citenamefont {Nakamura}, \citenamefont {Takada}, \citenamefont {Mogi}, \citenamefont {Takahashi}, \citenamefont {Tsukazaki}, \citenamefont {Kawasaki}, \citenamefont {Tokura},\ and\ \citenamefont {Kaneko}}]{okazaki_quantum_2022}%
  \BibitemOpen
  \bibfield  {author} {\bibinfo {author} {\bibfnamefont {Y.}~\bibnamefont {Okazaki}}, \bibinfo {author} {\bibfnamefont {T.}~\bibnamefont {Oe}}, \bibinfo {author} {\bibfnamefont {M.}~\bibnamefont {Kawamura}}, \bibinfo {author} {\bibfnamefont {R.}~\bibnamefont {Yoshimi}}, \bibinfo {author} {\bibfnamefont {S.}~\bibnamefont {Nakamura}}, \bibinfo {author} {\bibfnamefont {S.}~\bibnamefont {Takada}}, \bibinfo {author} {\bibfnamefont {M.}~\bibnamefont {Mogi}}, \bibinfo {author} {\bibfnamefont {K.~S.}\ \bibnamefont {Takahashi}}, \bibinfo {author} {\bibfnamefont {A.}~\bibnamefont {Tsukazaki}}, \bibinfo {author} {\bibfnamefont {M.}~\bibnamefont {Kawasaki}}, \bibinfo {author} {\bibfnamefont {Y.}~\bibnamefont {Tokura}},\ and\ \bibinfo {author} {\bibfnamefont {N.-H.}\ \bibnamefont {Kaneko}},\ }\bibfield  {title} {\bibinfo {title} {Quantum {Anomalous} {Hall} {Effect} with a {Permanent} {Magnet} {Defines} a {Quantum} {Resistance} {Standard}},\ }\href {https://doi.org/10.1038/s41567-021-01424-8} {\bibfield  {journal} {\bibinfo  {journal} {Nat. Phys.}\ }\textbf {\bibinfo {volume} {18}},\ \bibinfo {pages} {25} (\bibinfo {year} {2022})}\BibitemShut {NoStop}%
\bibitem [{\citenamefont {Chen}\ \emph {et~al.}(2019{\natexlab{a}})\citenamefont {Chen}, \citenamefont {Xu}, \citenamefont {Li}, \citenamefont {Li}, \citenamefont {Wang}, \citenamefont {Zhang}, \citenamefont {Li}, \citenamefont {Wu}, \citenamefont {Liang}, \citenamefont {Chen}, \citenamefont {Jung}, \citenamefont {Cacho}, \citenamefont {Mao}, \citenamefont {Liu}, \citenamefont {Wang}, \citenamefont {Guo}, \citenamefont {Xu}, \citenamefont {Liu}, \citenamefont {Yang},\ and\ \citenamefont {Chen}}]{chen_topological_2019}%
  \BibitemOpen
  \bibfield  {author} {\bibinfo {author} {\bibfnamefont {Y.~J.}\ \bibnamefont {Chen}}, \bibinfo {author} {\bibfnamefont {L.~X.}\ \bibnamefont {Xu}}, \bibinfo {author} {\bibfnamefont {J.~H.}\ \bibnamefont {Li}}, \bibinfo {author} {\bibfnamefont {Y.~W.}\ \bibnamefont {Li}}, \bibinfo {author} {\bibfnamefont {H.~Y.}\ \bibnamefont {Wang}}, \bibinfo {author} {\bibfnamefont {C.~F.}\ \bibnamefont {Zhang}}, \bibinfo {author} {\bibfnamefont {H.}~\bibnamefont {Li}}, \bibinfo {author} {\bibfnamefont {Y.}~\bibnamefont {Wu}}, \bibinfo {author} {\bibfnamefont {A.~J.}\ \bibnamefont {Liang}}, \bibinfo {author} {\bibfnamefont {C.}~\bibnamefont {Chen}}, \bibinfo {author} {\bibfnamefont {S.~W.}\ \bibnamefont {Jung}}, \bibinfo {author} {\bibfnamefont {C.}~\bibnamefont {Cacho}}, \bibinfo {author} {\bibfnamefont {Y.~H.}\ \bibnamefont {Mao}}, \bibinfo {author} {\bibfnamefont {S.}~\bibnamefont {Liu}}, \bibinfo {author} {\bibfnamefont {M.~X.}\ \bibnamefont {Wang}}, \bibinfo {author} {\bibfnamefont {Y.~F.}\ \bibnamefont {Guo}}, \bibinfo {author} {\bibfnamefont {Y.}~\bibnamefont {Xu}}, \bibinfo {author} {\bibfnamefont {Z.~K.}\ \bibnamefont {Liu}}, \bibinfo {author} {\bibfnamefont {L.~X.}\ \bibnamefont {Yang}},\ and\ \bibinfo {author} {\bibfnamefont {Y.~L.}\ \bibnamefont {Chen}},\ }\bibfield  {title} {\bibinfo {title} {Topological electronic structure and its temperature evolution in antiferromagnetic topological insulator mnbi2te4},\ }\href {https://doi.org/10.1103/PhysRevX.9.041040} {\bibfield  {journal} {\bibinfo  {journal} {Phys. Rev. X}\ }\textbf {\bibinfo {volume} {9}},\ \bibinfo {pages} {041040} (\bibinfo {year} {2019}{\natexlab{a}})}\BibitemShut {NoStop}%
\bibitem [{\citenamefont {Chen}\ \emph {et~al.}(2019{\natexlab{b}})\citenamefont {Chen}, \citenamefont {Fei}, \citenamefont {Zhang}, \citenamefont {Zhang}, \citenamefont {Liu}, \citenamefont {Zhang}, \citenamefont {Wang}, \citenamefont {Wei}, \citenamefont {Zhang}, \citenamefont {Zuo}, \citenamefont {Guo}, \citenamefont {Liu}, \citenamefont {Wang}, \citenamefont {Wu}, \citenamefont {Zong}, \citenamefont {Xie}, \citenamefont {Chen}, \citenamefont {Sun}, \citenamefont {Wang}, \citenamefont {Zhang}, \citenamefont {Zhang}, \citenamefont {Wang}, \citenamefont {Song}, \citenamefont {Zhang}, \citenamefont {Shen},\ and\ \citenamefont {Wang}}]{chen_intrinsic_2019}%
  \BibitemOpen
  \bibfield  {author} {\bibinfo {author} {\bibfnamefont {B.}~\bibnamefont {Chen}}, \bibinfo {author} {\bibfnamefont {F.}~\bibnamefont {Fei}}, \bibinfo {author} {\bibfnamefont {D.}~\bibnamefont {Zhang}}, \bibinfo {author} {\bibfnamefont {B.}~\bibnamefont {Zhang}}, \bibinfo {author} {\bibfnamefont {W.}~\bibnamefont {Liu}}, \bibinfo {author} {\bibfnamefont {S.}~\bibnamefont {Zhang}}, \bibinfo {author} {\bibfnamefont {P.}~\bibnamefont {Wang}}, \bibinfo {author} {\bibfnamefont {B.}~\bibnamefont {Wei}}, \bibinfo {author} {\bibfnamefont {Y.}~\bibnamefont {Zhang}}, \bibinfo {author} {\bibfnamefont {Z.}~\bibnamefont {Zuo}}, \bibinfo {author} {\bibfnamefont {J.}~\bibnamefont {Guo}}, \bibinfo {author} {\bibfnamefont {Q.}~\bibnamefont {Liu}}, \bibinfo {author} {\bibfnamefont {Z.}~\bibnamefont {Wang}}, \bibinfo {author} {\bibfnamefont {X.}~\bibnamefont {Wu}}, \bibinfo {author} {\bibfnamefont {J.}~\bibnamefont {Zong}}, \bibinfo {author} {\bibfnamefont {X.}~\bibnamefont {Xie}}, \bibinfo {author} {\bibfnamefont {W.}~\bibnamefont {Chen}}, \bibinfo {author} {\bibfnamefont {Z.}~\bibnamefont {Sun}}, \bibinfo {author} {\bibfnamefont {S.}~\bibnamefont {Wang}}, \bibinfo {author} {\bibfnamefont {Y.}~\bibnamefont {Zhang}}, \bibinfo {author} {\bibfnamefont {M.}~\bibnamefont {Zhang}}, \bibinfo {author} {\bibfnamefont {X.}~\bibnamefont {Wang}}, \bibinfo {author} {\bibfnamefont {F.}~\bibnamefont {Song}}, \bibinfo {author} {\bibfnamefont {H.}~\bibnamefont {Zhang}}, \bibinfo {author} {\bibfnamefont {D.}~\bibnamefont {Shen}},\ and\ \bibinfo {author} {\bibfnamefont {B.}~\bibnamefont {Wang}},\ }\bibfield  {title} {\bibinfo {title} {Intrinsic {Magnetic} {Topological} {Insulator} {Phases} in the {Sb} {Doped} {MnBi2Te4} {Bulks} and {Thin} {Flakes}},\ }\href {https://doi.org/10.1038/s41467-019-12485-y} {\bibfield  {journal} {\bibinfo  {journal} {Nat Commun}\ }\textbf {\bibinfo {volume} {10}},\ \bibinfo {pages} {4469} (\bibinfo {year} {2019}{\natexlab{b}})}\BibitemShut {NoStop}%
\bibitem [{\citenamefont {Liu}\ \emph {et~al.}(2020)\citenamefont {Liu}, \citenamefont {Wang}, \citenamefont {Li}, \citenamefont {Wu}, \citenamefont {Li}, \citenamefont {Li}, \citenamefont {He}, \citenamefont {Xu}, \citenamefont {Zhang},\ and\ \citenamefont {Wang}}]{liu_robust_2020}%
  \BibitemOpen
  \bibfield  {author} {\bibinfo {author} {\bibfnamefont {C.}~\bibnamefont {Liu}}, \bibinfo {author} {\bibfnamefont {Y.}~\bibnamefont {Wang}}, \bibinfo {author} {\bibfnamefont {H.}~\bibnamefont {Li}}, \bibinfo {author} {\bibfnamefont {Y.}~\bibnamefont {Wu}}, \bibinfo {author} {\bibfnamefont {Y.}~\bibnamefont {Li}}, \bibinfo {author} {\bibfnamefont {J.}~\bibnamefont {Li}}, \bibinfo {author} {\bibfnamefont {K.}~\bibnamefont {He}}, \bibinfo {author} {\bibfnamefont {Y.}~\bibnamefont {Xu}}, \bibinfo {author} {\bibfnamefont {J.}~\bibnamefont {Zhang}},\ and\ \bibinfo {author} {\bibfnamefont {Y.}~\bibnamefont {Wang}},\ }\bibfield  {title} {\bibinfo {title} {Robust {Axion} {Insulator} and {Chern} {Insulator} {Phases} in a {Two}-{Dimensional} {Antiferromagnetic} {Topological} {Insulator}},\ }\href {https://doi.org/10.1038/s41563-019-0573-3} {\bibfield  {journal} {\bibinfo  {journal} {Nat. Mater.}\ }\textbf {\bibinfo {volume} {19}},\ \bibinfo {pages} {522} (\bibinfo {year} {2020})}\BibitemShut {NoStop}%
\bibitem [{\citenamefont {Liang}\ \emph {et~al.}(2025)\citenamefont {Liang}, \citenamefont {Li}, \citenamefont {An}, \citenamefont {Ren}, \citenamefont {Qiao},\ and\ \citenamefont {Niu}}]{liang_chern_2025}%
  \BibitemOpen
  \bibfield  {author} {\bibinfo {author} {\bibfnamefont {W.}~\bibnamefont {Liang}}, \bibinfo {author} {\bibfnamefont {Z.}~\bibnamefont {Li}}, \bibinfo {author} {\bibfnamefont {J.}~\bibnamefont {An}}, \bibinfo {author} {\bibfnamefont {Y.}~\bibnamefont {Ren}}, \bibinfo {author} {\bibfnamefont {Z.}~\bibnamefont {Qiao}},\ and\ \bibinfo {author} {\bibfnamefont {Q.}~\bibnamefont {Niu}},\ }\bibfield  {title} {\bibinfo {title} {Chern {Number} {Tunable} {Quantum} {Anomalous} {Hall} {Effect} in {Compensated} {Antiferromagnets}},\ }\href {https://doi.org/10.1103/PhysRevLett.134.116603} {\bibfield  {journal} {\bibinfo  {journal} {Phys. Rev. Lett.}\ }\textbf {\bibinfo {volume} {134}},\ \bibinfo {pages} {116603} (\bibinfo {year} {2025})}\BibitemShut {NoStop}%
\bibitem [{\citenamefont {Bernevig}\ and\ \citenamefont {Hughes}(2013)}]{bernevig_topological_2013}%
  \BibitemOpen
  \bibfield  {author} {\bibinfo {author} {\bibfnamefont {B.~A.}\ \bibnamefont {Bernevig}}\ and\ \bibinfo {author} {\bibfnamefont {T.~L.}\ \bibnamefont {Hughes}},\ }\href@noop {} {\emph {\bibinfo {title} {Topological {Insulators} and {Topological} {Superconductors}}}}\ (\bibinfo  {publisher} {Princeton University Press},\ \bibinfo {address} {Princeton, New Jersey},\ \bibinfo {year} {2013})\BibitemShut {NoStop}%
\bibitem [{\citenamefont {Lee}\ \emph {et~al.}(2010)\citenamefont {Lee}, \citenamefont {Fang}, \citenamefont {Vlahos}, \citenamefont {Ke}, \citenamefont {Jung}, \citenamefont {Kourkoutis}, \citenamefont {Kim}, \citenamefont {Ryan}, \citenamefont {Heeg}, \citenamefont {Roeckerath}, \citenamefont {Goian}, \citenamefont {Bernhagen}, \citenamefont {Uecker}, \citenamefont {Hammel}, \citenamefont {Rabe}, \citenamefont {Kamba}, \citenamefont {Schubert}, \citenamefont {Freeland}, \citenamefont {Muller}, \citenamefont {Fennie}, \citenamefont {Schiffer}, \citenamefont {Gopalan}, \citenamefont {Johnston-Halperin},\ and\ \citenamefont {Schlom}}]{lee_strong_2010}%
  \BibitemOpen
  \bibfield  {author} {\bibinfo {author} {\bibfnamefont {J.~H.}\ \bibnamefont {Lee}}, \bibinfo {author} {\bibfnamefont {L.}~\bibnamefont {Fang}}, \bibinfo {author} {\bibfnamefont {E.}~\bibnamefont {Vlahos}}, \bibinfo {author} {\bibfnamefont {X.}~\bibnamefont {Ke}}, \bibinfo {author} {\bibfnamefont {Y.~W.}\ \bibnamefont {Jung}}, \bibinfo {author} {\bibfnamefont {L.~F.}\ \bibnamefont {Kourkoutis}}, \bibinfo {author} {\bibfnamefont {J.-W.}\ \bibnamefont {Kim}}, \bibinfo {author} {\bibfnamefont {P.~J.}\ \bibnamefont {Ryan}}, \bibinfo {author} {\bibfnamefont {T.}~\bibnamefont {Heeg}}, \bibinfo {author} {\bibfnamefont {M.}~\bibnamefont {Roeckerath}}, \bibinfo {author} {\bibfnamefont {V.}~\bibnamefont {Goian}}, \bibinfo {author} {\bibfnamefont {M.}~\bibnamefont {Bernhagen}}, \bibinfo {author} {\bibfnamefont {R.}~\bibnamefont {Uecker}}, \bibinfo {author} {\bibfnamefont {P.~C.}\ \bibnamefont {Hammel}}, \bibinfo {author} {\bibfnamefont {K.~M.}\ \bibnamefont {Rabe}}, \bibinfo {author} {\bibfnamefont {S.}~\bibnamefont {Kamba}}, \bibinfo {author} {\bibfnamefont {J.}~\bibnamefont {Schubert}}, \bibinfo {author} {\bibfnamefont {J.~W.}\ \bibnamefont {Freeland}}, \bibinfo {author} {\bibfnamefont {D.~A.}\ \bibnamefont {Muller}}, \bibinfo {author} {\bibfnamefont {C.~J.}\ \bibnamefont {Fennie}}, \bibinfo {author} {\bibfnamefont {P.}~\bibnamefont {Schiffer}}, \bibinfo {author} {\bibfnamefont {V.}~\bibnamefont {Gopalan}}, \bibinfo {author} {\bibfnamefont {E.}~\bibnamefont {Johnston-Halperin}},\ and\ \bibinfo {author} {\bibfnamefont {D.~G.}\ \bibnamefont {Schlom}},\ }\bibfield  {title} {\bibinfo {title} {A strong ferroelectric ferromagnet created by means of spin-lattice coupling},\ }\href {https://doi.org/10.1038/nature09331} {\bibfield  {journal} {\bibinfo  {journal} {Nature}\ }\textbf {\bibinfo {volume} {466}},\ \bibinfo {pages} {954} (\bibinfo {year} {2010})}\BibitemShut {NoStop}%
\bibitem [{\citenamefont {Bai}\ \emph {et~al.}(2023)\citenamefont {Bai}, \citenamefont {Fu}, \citenamefont {Zhang},\ and\ \citenamefont {Shen}}]{bai_metallic_2023}%
  \BibitemOpen
  \bibfield  {author} {\bibinfo {author} {\bibfnamefont {K.-Z.}\ \bibnamefont {Bai}}, \bibinfo {author} {\bibfnamefont {B.}~\bibnamefont {Fu}}, \bibinfo {author} {\bibfnamefont {Z.}~\bibnamefont {Zhang}},\ and\ \bibinfo {author} {\bibfnamefont {S.-Q.}\ \bibnamefont {Shen}},\ }\bibfield  {title} {\bibinfo {title} {Metallic {Quantized} {Anomalous} {Hall} {Effect} without {Chiral} {Edge} {States}},\ }\href {https://doi.org/10.1103/PhysRevB.108.L241407} {\bibfield  {journal} {\bibinfo  {journal} {Phys. Rev. B}\ }\textbf {\bibinfo {volume} {108}},\ \bibinfo {pages} {L241407} (\bibinfo {year} {2023})}\BibitemShut {NoStop}%
\bibitem [{\citenamefont {Mogi}\ \emph {et~al.}(2022)\citenamefont {Mogi}, \citenamefont {Okamura}, \citenamefont {Kawamura}, \citenamefont {Yoshimi}, \citenamefont {Yasuda}, \citenamefont {Tsukazaki}, \citenamefont {Takahashi}, \citenamefont {Morimoto}, \citenamefont {Nagaosa}, \citenamefont {Kawasaki}, \citenamefont {Takahashi},\ and\ \citenamefont {Tokura}}]{mogi_experimental_2022}%
  \BibitemOpen
  \bibfield  {author} {\bibinfo {author} {\bibfnamefont {M.}~\bibnamefont {Mogi}}, \bibinfo {author} {\bibfnamefont {Y.}~\bibnamefont {Okamura}}, \bibinfo {author} {\bibfnamefont {M.}~\bibnamefont {Kawamura}}, \bibinfo {author} {\bibfnamefont {R.}~\bibnamefont {Yoshimi}}, \bibinfo {author} {\bibfnamefont {K.}~\bibnamefont {Yasuda}}, \bibinfo {author} {\bibfnamefont {A.}~\bibnamefont {Tsukazaki}}, \bibinfo {author} {\bibfnamefont {K.~S.}\ \bibnamefont {Takahashi}}, \bibinfo {author} {\bibfnamefont {T.}~\bibnamefont {Morimoto}}, \bibinfo {author} {\bibfnamefont {N.}~\bibnamefont {Nagaosa}}, \bibinfo {author} {\bibfnamefont {M.}~\bibnamefont {Kawasaki}}, \bibinfo {author} {\bibfnamefont {Y.}~\bibnamefont {Takahashi}},\ and\ \bibinfo {author} {\bibfnamefont {Y.}~\bibnamefont {Tokura}},\ }\bibfield  {title} {\bibinfo {title} {Experimental signature of the parity anomaly in a semi-magnetic topological insulator},\ }\href {https://doi.org/10.1038/s41567-021-01490-y} {\bibfield  {journal} {\bibinfo  {journal} {Nat. Phys.}\ }\textbf {\bibinfo {volume} {18}},\ \bibinfo {pages} {390} (\bibinfo {year} {2022})}\BibitemShut {NoStop}%
\bibitem [{\citenamefont {Fu}\ \emph {et~al.}(2022)\citenamefont {Fu}, \citenamefont {Zou}, \citenamefont {Hu}, \citenamefont {Wang},\ and\ \citenamefont {Shen}}]{fu2022quantum}%
  \BibitemOpen
  \bibfield  {author} {\bibinfo {author} {\bibfnamefont {B.}~\bibnamefont {Fu}}, \bibinfo {author} {\bibfnamefont {J.-Y.}\ \bibnamefont {Zou}}, \bibinfo {author} {\bibfnamefont {Z.-A.}\ \bibnamefont {Hu}}, \bibinfo {author} {\bibfnamefont {H.-W.}\ \bibnamefont {Wang}},\ and\ \bibinfo {author} {\bibfnamefont {S.-Q.}\ \bibnamefont {Shen}},\ }\bibfield  {title} {\bibinfo {title} {Quantum anomalous semimetals},\ }\href@noop {} {\bibfield  {journal} {\bibinfo  {journal} {npj Quantum Materials}\ }\textbf {\bibinfo {volume} {7}},\ \bibinfo {pages} {94} (\bibinfo {year} {2022})}\BibitemShut {NoStop}%
\bibitem [{\citenamefont {Zou}\ \emph {et~al.}(2022)\citenamefont {Zou}, \citenamefont {Fu}, \citenamefont {Wang}, \citenamefont {Hu},\ and\ \citenamefont {Shen}}]{zou2022half}%
  \BibitemOpen
  \bibfield  {author} {\bibinfo {author} {\bibfnamefont {J.-Y.}\ \bibnamefont {Zou}}, \bibinfo {author} {\bibfnamefont {B.}~\bibnamefont {Fu}}, \bibinfo {author} {\bibfnamefont {H.-W.}\ \bibnamefont {Wang}}, \bibinfo {author} {\bibfnamefont {Z.-A.}\ \bibnamefont {Hu}},\ and\ \bibinfo {author} {\bibfnamefont {S.-Q.}\ \bibnamefont {Shen}},\ }\bibfield  {title} {\bibinfo {title} {Half-quantized hall effect and power law decay of edge-current distribution},\ }\href@noop {} {\bibfield  {journal} {\bibinfo  {journal} {Physical Review B}\ }\textbf {\bibinfo {volume} {105}},\ \bibinfo {pages} {L201106} (\bibinfo {year} {2022})}\BibitemShut {NoStop}%
\bibitem [{\citenamefont {Zou}\ \emph {et~al.}(2023)\citenamefont {Zou}, \citenamefont {Chen}, \citenamefont {Fu}, \citenamefont {Wang}, \citenamefont {Hu},\ and\ \citenamefont {Shen}}]{zou2023half}%
  \BibitemOpen
  \bibfield  {author} {\bibinfo {author} {\bibfnamefont {J.-Y.}\ \bibnamefont {Zou}}, \bibinfo {author} {\bibfnamefont {R.}~\bibnamefont {Chen}}, \bibinfo {author} {\bibfnamefont {B.}~\bibnamefont {Fu}}, \bibinfo {author} {\bibfnamefont {H.-W.}\ \bibnamefont {Wang}}, \bibinfo {author} {\bibfnamefont {Z.-A.}\ \bibnamefont {Hu}},\ and\ \bibinfo {author} {\bibfnamefont {S.-Q.}\ \bibnamefont {Shen}},\ }\bibfield  {title} {\bibinfo {title} {Half-quantized hall effect at the parity-invariant fermi surface},\ }\href@noop {} {\bibfield  {journal} {\bibinfo  {journal} {Physical Review B}\ }\textbf {\bibinfo {volume} {107}},\ \bibinfo {pages} {125153} (\bibinfo {year} {2023})}\BibitemShut {NoStop}%
\bibitem [{\citenamefont {Ning}\ \emph {et~al.}(2023)\citenamefont {Ning}, \citenamefont {Ding}, \citenamefont {Xu},\ and\ \citenamefont {Wang}}]{ning2023robustness}%
  \BibitemOpen
  \bibfield  {author} {\bibinfo {author} {\bibfnamefont {Z.}~\bibnamefont {Ning}}, \bibinfo {author} {\bibfnamefont {X.}~\bibnamefont {Ding}}, \bibinfo {author} {\bibfnamefont {D.-H.}\ \bibnamefont {Xu}},\ and\ \bibinfo {author} {\bibfnamefont {R.}~\bibnamefont {Wang}},\ }\bibfield  {title} {\bibinfo {title} {Robustness of half-integer quantized hall conductivity against disorder in an anisotropic dirac semimetal with parity anomaly},\ }\href@noop {} {\bibfield  {journal} {\bibinfo  {journal} {Physical Review B}\ }\textbf {\bibinfo {volume} {108}},\ \bibinfo {pages} {L041104} (\bibinfo {year} {2023})}\BibitemShut {NoStop}%
\bibitem [{\citenamefont {Wang}\ \emph {et~al.}(2024)\citenamefont {Wang}, \citenamefont {Fu},\ and\ \citenamefont {Shen}}]{wang2024signature}%
  \BibitemOpen
  \bibfield  {author} {\bibinfo {author} {\bibfnamefont {H.-W.}\ \bibnamefont {Wang}}, \bibinfo {author} {\bibfnamefont {B.}~\bibnamefont {Fu}},\ and\ \bibinfo {author} {\bibfnamefont {S.-Q.}\ \bibnamefont {Shen}},\ }\bibfield  {title} {\bibinfo {title} {Signature of parity anomaly: Crossover from one half to integer quantized hall conductance in a finite magnetic field},\ }\href@noop {} {\bibfield  {journal} {\bibinfo  {journal} {Physical Review B}\ }\textbf {\bibinfo {volume} {109}},\ \bibinfo {pages} {075113} (\bibinfo {year} {2024})}\BibitemShut {NoStop}%
\bibitem [{\citenamefont {Wan}\ and\ \citenamefont {Sun}(2024{\natexlab{a}})}]{wan_quarter-quantized_2024}%
  \BibitemOpen
  \bibfield  {author} {\bibinfo {author} {\bibfnamefont {Y.-H.}\ \bibnamefont {Wan}}\ and\ \bibinfo {author} {\bibfnamefont {Q.-F.}\ \bibnamefont {Sun}},\ }\bibfield  {title} {\bibinfo {title} {Quarter-{Quantized} {Thermal} {Hall} {Effect} with {Parity} {Anomaly}},\ }\href {https://doi.org/10.1103/PhysRevB.109.195408} {\bibfield  {journal} {\bibinfo  {journal} {Phys. Rev. B}\ }\textbf {\bibinfo {volume} {109}},\ \bibinfo {pages} {195408} (\bibinfo {year} {2024}{\natexlab{a}})}\BibitemShut {NoStop}%
\bibitem [{\citenamefont {Fu}\ \emph {et~al.}(2024)\citenamefont {Fu}, \citenamefont {Bai},\ and\ \citenamefont {Shen}}]{fu2024half}%
  \BibitemOpen
  \bibfield  {author} {\bibinfo {author} {\bibfnamefont {B.}~\bibnamefont {Fu}}, \bibinfo {author} {\bibfnamefont {K.-Z.}\ \bibnamefont {Bai}},\ and\ \bibinfo {author} {\bibfnamefont {S.-Q.}\ \bibnamefont {Shen}},\ }\bibfield  {title} {\bibinfo {title} {Half-quantum mirror hall effect},\ }\href@noop {} {\bibfield  {journal} {\bibinfo  {journal} {Nature Communications}\ }\textbf {\bibinfo {volume} {15}},\ \bibinfo {pages} {6939} (\bibinfo {year} {2024})}\BibitemShut {NoStop}%
\bibitem [{\citenamefont {Nagaosa}\ \emph {et~al.}(2010)\citenamefont {Nagaosa}, \citenamefont {Sinova}, \citenamefont {Onoda}, \citenamefont {MacDonald},\ and\ \citenamefont {Ong}}]{nagaosa_anomalous_2010}%
  \BibitemOpen
  \bibfield  {author} {\bibinfo {author} {\bibfnamefont {N.}~\bibnamefont {Nagaosa}}, \bibinfo {author} {\bibfnamefont {J.}~\bibnamefont {Sinova}}, \bibinfo {author} {\bibfnamefont {S.}~\bibnamefont {Onoda}}, \bibinfo {author} {\bibfnamefont {A.~H.}\ \bibnamefont {MacDonald}},\ and\ \bibinfo {author} {\bibfnamefont {N.~P.}\ \bibnamefont {Ong}},\ }\bibfield  {title} {\bibinfo {title} {Anomalous {Hall} effect},\ }\href {https://doi.org/10.1103/RevModPhys.82.1539} {\bibfield  {journal} {\bibinfo  {journal} {Rev. Mod. Phys.}\ }\textbf {\bibinfo {volume} {82}},\ \bibinfo {pages} {1539} (\bibinfo {year} {2010})}\BibitemShut {NoStop}%
\bibitem [{see()}]{seesupplemental}%
  \BibitemOpen
  \bibinfo {title} {See supplemental material at [url will be inserted by publisher] for detail.}\BibitemShut {Stop}%
\bibitem [{\citenamefont {Qi}\ \emph {et~al.}(2006)\citenamefont {Qi}, \citenamefont {Wu},\ and\ \citenamefont {Zhang}}]{qi_topological_2006}%
  \BibitemOpen
\bibfield  {title} {  }\bibfield  {author} {\bibinfo {author} {\bibfnamefont {X.-L.}\ \bibnamefont {Qi}}, \bibinfo {author} {\bibfnamefont {Y.-S.}\ \bibnamefont {Wu}},\ and\ \bibinfo {author} {\bibfnamefont {S.-C.}\ \bibnamefont {Zhang}},\ }\bibfield  {title} {\bibinfo {title} {Topological quantization of the spin {Hall} effect in two-dimensional paramagnetic semiconductors},\ }\href {https://doi.org/10.1103/PhysRevB.74.085308} {\bibfield  {journal} {\bibinfo  {journal} {Phys. Rev. B}\ }\textbf {\bibinfo {volume} {74}},\ \bibinfo {pages} {085308} (\bibinfo {year} {2006})}\BibitemShut {NoStop}%
\bibitem [{\citenamefont {Zhou}\ \emph {et~al.}(2022)\citenamefont {Zhou}, \citenamefont {Li}, \citenamefont {Xu}, \citenamefont {Chen}, \citenamefont {Sun},\ and\ \citenamefont {Xie}}]{zhou_transport_2022}%
  \BibitemOpen
  \bibfield  {author} {\bibinfo {author} {\bibfnamefont {H.}~\bibnamefont {Zhou}}, \bibinfo {author} {\bibfnamefont {H.}~\bibnamefont {Li}}, \bibinfo {author} {\bibfnamefont {D.-H.}\ \bibnamefont {Xu}}, \bibinfo {author} {\bibfnamefont {C.-Z.}\ \bibnamefont {Chen}}, \bibinfo {author} {\bibfnamefont {Q.-F.}\ \bibnamefont {Sun}},\ and\ \bibinfo {author} {\bibfnamefont {X.~C.}\ \bibnamefont {Xie}},\ }\bibfield  {title} {\bibinfo {title} {Transport {Theory} of {Half}-{Quantized} {Hall} {Conductance} in a {Semimagnetic} {Topological} {Insulator}},\ }\href {https://doi.org/10.1103/PhysRevLett.129.096601} {\bibfield  {journal} {\bibinfo  {journal} {Phys. Rev. Lett.}\ }\textbf {\bibinfo {volume} {129}},\ \bibinfo {pages} {096601} (\bibinfo {year} {2022})}\BibitemShut {NoStop}%
\bibitem [{\citenamefont {B\"{u}ttiker}(1986)}]{buttiker_four-terminal_1986}%
  \BibitemOpen
  \bibfield  {author} {\bibinfo {author} {\bibfnamefont {M.}~\bibnamefont {B\"{u}ttiker}},\ }\bibfield  {title} {\bibinfo {title} {Four-{Terminal} {Phase}-{Coherent} {Conductance}},\ }\href {https://doi.org/10.1103/PhysRevLett.57.1761} {\bibfield  {journal} {\bibinfo  {journal} {Phys. Rev. Lett.}\ }\textbf {\bibinfo {volume} {57}},\ \bibinfo {pages} {1761} (\bibinfo {year} {1986})}\BibitemShut {NoStop}%
\bibitem [{\citenamefont {B\"{u}ttiker}(1988{\natexlab{a}})}]{buttiker_symmetry_1988}%
  \BibitemOpen
  \bibfield  {author} {\bibinfo {author} {\bibfnamefont {M.}~\bibnamefont {B\"{u}ttiker}},\ }\bibfield  {title} {\bibinfo {title} {Symmetry of {Electrical} {Conduction}},\ }\href {https://doi.org/10.1147/rd.323.0317} {\bibfield  {journal} {\bibinfo  {journal} {IBM Journal of Research and Development}\ }\textbf {\bibinfo {volume} {32}},\ \bibinfo {pages} {317} (\bibinfo {year} {1988}{\natexlab{a}})}\BibitemShut {NoStop}%
\bibitem [{\citenamefont {B\"{u}ttiker}(1988{\natexlab{b}})}]{buttiker_absence_1988}%
  \BibitemOpen
  \bibfield  {author} {\bibinfo {author} {\bibfnamefont {M.}~\bibnamefont {B\"{u}ttiker}},\ }\bibfield  {title} {\bibinfo {title} {Absence of {Backscattering} in the {Quantum} {Hall} {Effect} in {Multiprobe} {Conductors}},\ }\href {https://doi.org/10.1103/PhysRevB.38.9375} {\bibfield  {journal} {\bibinfo  {journal} {Phys. Rev. B}\ }\textbf {\bibinfo {volume} {38}},\ \bibinfo {pages} {9375} (\bibinfo {year} {1988}{\natexlab{b}})}\BibitemShut {NoStop}%
\bibitem [{\citenamefont {Wan}\ and\ \citenamefont {Sun}(2025)}]{wan_altermagnetism-induced_2025}%
  \BibitemOpen
  \bibfield  {author} {\bibinfo {author} {\bibfnamefont {Y.-H.}\ \bibnamefont {Wan}}\ and\ \bibinfo {author} {\bibfnamefont {Q.-F.}\ \bibnamefont {Sun}},\ }\bibfield  {title} {\bibinfo {title} {Altermagnetism-{Induced} {Parity} {Anomaly} in {Weak} {Topological} {Insulators}},\ }\href {https://doi.org/10.1103/PhysRevB.111.045407} {\bibfield  {journal} {\bibinfo  {journal} {Phys. Rev. B}\ }\textbf {\bibinfo {volume} {111}},\ \bibinfo {pages} {045407} (\bibinfo {year} {2025})}\BibitemShut {NoStop}%
\bibitem [{\citenamefont {McLennan}\ \emph {et~al.}(1991)\citenamefont {McLennan}, \citenamefont {Lee},\ and\ \citenamefont {Datta}}]{mclennan_voltage_1991}%
  \BibitemOpen
  \bibfield  {author} {\bibinfo {author} {\bibfnamefont {M.~J.}\ \bibnamefont {McLennan}}, \bibinfo {author} {\bibfnamefont {Y.}~\bibnamefont {Lee}},\ and\ \bibinfo {author} {\bibfnamefont {S.}~\bibnamefont {Datta}},\ }\bibfield  {title} {\bibinfo {title} {Voltage drop in mesoscopic systems: {A} numerical study using a quantum kinetic equation},\ }\href {https://doi.org/10.1103/PhysRevB.43.13846} {\bibfield  {journal} {\bibinfo  {journal} {Phys. Rev. B}\ }\textbf {\bibinfo {volume} {43}},\ \bibinfo {pages} {13846} (\bibinfo {year} {1991})}\BibitemShut {NoStop}%
\bibitem [{\citenamefont {Liu}\ and\ \citenamefont {Sun}(2024)}]{liu_dissipation_2024}%
  \BibitemOpen
  \bibfield  {author} {\bibinfo {author} {\bibfnamefont {P.-Y.}\ \bibnamefont {Liu}}\ and\ \bibinfo {author} {\bibfnamefont {Q.-F.}\ \bibnamefont {Sun}},\ }\bibfield  {title} {\bibinfo {title} {Dissipation and dephasing in quantum {Hall} interferometers},\ }\href {https://doi.org/10.1103/PhysRevB.110.085411} {\bibfield  {journal} {\bibinfo  {journal} {Phys. Rev. B}\ }\textbf {\bibinfo {volume} {110}},\ \bibinfo {pages} {085411} (\bibinfo {year} {2024})}\BibitemShut {NoStop}%
\bibitem [{\citenamefont {Meir}\ and\ \citenamefont {Wingreen}(1992)}]{meir_landauer_1992}%
  \BibitemOpen
  \bibfield  {author} {\bibinfo {author} {\bibfnamefont {Y.}~\bibnamefont {Meir}}\ and\ \bibinfo {author} {\bibfnamefont {N.~S.}\ \bibnamefont {Wingreen}},\ }\bibfield  {title} {\bibinfo {title} {Landauer formula for the current through an interacting electron region},\ }\href {https://doi.org/10.1103/PhysRevLett.68.2512} {\bibfield  {journal} {\bibinfo  {journal} {Phys. Rev. Lett.}\ }\textbf {\bibinfo {volume} {68}},\ \bibinfo {pages} {2512} (\bibinfo {year} {1992})}\BibitemShut {NoStop}%
\bibitem [{\citenamefont {Lambert}\ \emph {et~al.}(1993)\citenamefont {Lambert}, \citenamefont {Hui},\ and\ \citenamefont {Robinson}}]{lambert_multi-probe_1993}%
  \BibitemOpen
  \bibfield  {author} {\bibinfo {author} {\bibfnamefont {C.~J.}\ \bibnamefont {Lambert}}, \bibinfo {author} {\bibfnamefont {V.~C.}\ \bibnamefont {Hui}},\ and\ \bibinfo {author} {\bibfnamefont {S.~J.}\ \bibnamefont {Robinson}},\ }\bibfield  {title} {\bibinfo {title} {Multi-{Probe} {Conductance} {Formulae} for {Mesoscopic} {Superconductors}},\ }\href {https://doi.org/10.1088/0953-8984/5/25/009} {\bibfield  {journal} {\bibinfo  {journal} {J. Phys.: Condens. Matter}\ }\textbf {\bibinfo {volume} {5}},\ \bibinfo {pages} {4187} (\bibinfo {year} {1993})}\BibitemShut {NoStop}%
\bibitem [{\citenamefont {Wan}\ and\ \citenamefont {Sun}(2024{\natexlab{b}})}]{wan_mag}%
  \BibitemOpen
  \bibfield  {author} {\bibinfo {author} {\bibfnamefont {Y.-H.}\ \bibnamefont {Wan}}\ and\ \bibinfo {author} {\bibfnamefont {Q.-F.}\ \bibnamefont {Sun}},\ }\bibfield  {title} {\bibinfo {title} {Magnetization-induced phase transitions on the surface of three-dimensional topological insulators},\ }\href {https://doi.org/10.1103/PhysRevB.109.045418} {\bibfield  {journal} {\bibinfo  {journal} {Phys. Rev. B}\ }\textbf {\bibinfo {volume} {109}},\ \bibinfo {pages} {045418} (\bibinfo {year} {2024}{\natexlab{b}})}\BibitemShut {NoStop}%
\bibitem [{\citenamefont {Youn}\ \emph {et~al.}(2008)\citenamefont {Youn}, \citenamefont {Lee},\ and\ \citenamefont {Sim}}]{youn2008nonequilibrium}%
  \BibitemOpen
  \bibfield  {author} {\bibinfo {author} {\bibfnamefont {S.-C.}\ \bibnamefont {Youn}}, \bibinfo {author} {\bibfnamefont {H.-W.}\ \bibnamefont {Lee}},\ and\ \bibinfo {author} {\bibfnamefont {H.-S.}\ \bibnamefont {Sim}},\ }\bibfield  {title} {\bibinfo {title} {Nonequilibrium dephasing in an electronic mach-zehnder interferometer},\ }\href@noop {} {\bibfield  {journal} {\bibinfo  {journal} {Physical review letters}\ }\textbf {\bibinfo {volume} {100}},\ \bibinfo {pages} {196807} (\bibinfo {year} {2008})}\BibitemShut {NoStop}%
\bibitem [{\citenamefont {Chakravarty}\ and\ \citenamefont {Schmid}(1986)}]{chakravarty1986weak}%
  \BibitemOpen
  \bibfield  {author} {\bibinfo {author} {\bibfnamefont {S.}~\bibnamefont {Chakravarty}}\ and\ \bibinfo {author} {\bibfnamefont {A.}~\bibnamefont {Schmid}},\ }\bibfield  {title} {\bibinfo {title} {Weak localization: The quasiclassical theory of electrons in a random potential},\ }\href@noop {} {\bibfield  {journal} {\bibinfo  {journal} {Physics Reports}\ }\textbf {\bibinfo {volume} {140}},\ \bibinfo {pages} {193} (\bibinfo {year} {1986})}\BibitemShut {NoStop}%
\bibitem [{\citenamefont {Stern}\ \emph {et~al.}(1990)\citenamefont {Stern}, \citenamefont {Aharonov},\ and\ \citenamefont {Imry}}]{stern1990phase}%
  \BibitemOpen
  \bibfield  {author} {\bibinfo {author} {\bibfnamefont {A.}~\bibnamefont {Stern}}, \bibinfo {author} {\bibfnamefont {Y.}~\bibnamefont {Aharonov}},\ and\ \bibinfo {author} {\bibfnamefont {Y.}~\bibnamefont {Imry}},\ }\bibfield  {title} {\bibinfo {title} {Phase uncertainty and loss of interference: A general picture},\ }\href@noop {} {\bibfield  {journal} {\bibinfo  {journal} {Physical Review A}\ }\textbf {\bibinfo {volume} {41}},\ \bibinfo {pages} {3436} (\bibinfo {year} {1990})}\BibitemShut {NoStop}%
\bibitem [{\citenamefont {Burkard}\ \emph {et~al.}(1999)\citenamefont {Burkard}, \citenamefont {Loss},\ and\ \citenamefont {DiVincenzo}}]{burkard1999coupled}%
  \BibitemOpen
  \bibfield  {author} {\bibinfo {author} {\bibfnamefont {G.}~\bibnamefont {Burkard}}, \bibinfo {author} {\bibfnamefont {D.}~\bibnamefont {Loss}},\ and\ \bibinfo {author} {\bibfnamefont {D.~P.}\ \bibnamefont {DiVincenzo}},\ }\bibfield  {title} {\bibinfo {title} {Coupled quantum dots as quantum gates},\ }\href@noop {} {\bibfield  {journal} {\bibinfo  {journal} {Physical Review B}\ }\textbf {\bibinfo {volume} {59}},\ \bibinfo {pages} {2070} (\bibinfo {year} {1999})}\BibitemShut {NoStop}%
\bibitem [{\citenamefont {Fang}\ \emph {et~al.}(2023)\citenamefont {Fang}, \citenamefont {Guo},\ and\ \citenamefont {Sun}}]{addr1}%
  \BibitemOpen
  \bibfield  {author} {\bibinfo {author} {\bibfnamefont {J.-Y.}\ \bibnamefont {Fang}}, \bibinfo {author} {\bibfnamefont {A.-M.}\ \bibnamefont {Guo}},\ and\ \bibinfo {author} {\bibfnamefont {Q.-F.}\ \bibnamefont {Sun}},\ }\bibfield  {title} {\bibinfo {title} {Dephasing effect promotes the appearance of quantized hall plateaus},\ }\href {https://doi.org/10.1088/1367-2630/acbed2} {\bibfield  {journal} {\bibinfo  {journal} {New Journal of Physics}\ }\textbf {\bibinfo {volume} {25}},\ \bibinfo {pages} {033001} (\bibinfo {year} {2023})}\BibitemShut {NoStop}%
\bibitem [{\citenamefont {Huckestein}(1995)}]{huckestein_scaling_1995}%
  \BibitemOpen
  \bibfield  {author} {\bibinfo {author} {\bibfnamefont {B.}~\bibnamefont {Huckestein}},\ }\bibfield  {title} {\bibinfo {title} {Scaling theory of the integer quantum {Hall} effect},\ }\href {https://doi.org/10.1103/RevModPhys.67.357} {\bibfield  {journal} {\bibinfo  {journal} {Rev. Mod. Phys.}\ }\textbf {\bibinfo {volume} {67}},\ \bibinfo {pages} {357} (\bibinfo {year} {1995})}\BibitemShut {NoStop}%
\bibitem [{\citenamefont {Deng}\ \emph {et~al.}(2022)\citenamefont {Deng}, \citenamefont {Eckberg}, \citenamefont {Zhang}, \citenamefont {Qiu}, \citenamefont {Emmanouilidou}, \citenamefont {Yin}, \citenamefont {Chong}, \citenamefont {Tai}, \citenamefont {Ni},\ and\ \citenamefont {Wang}}]{deng_probing_2022}%
  \BibitemOpen
  \bibfield  {author} {\bibinfo {author} {\bibfnamefont {P.}~\bibnamefont {Deng}}, \bibinfo {author} {\bibfnamefont {C.}~\bibnamefont {Eckberg}}, \bibinfo {author} {\bibfnamefont {P.}~\bibnamefont {Zhang}}, \bibinfo {author} {\bibfnamefont {G.}~\bibnamefont {Qiu}}, \bibinfo {author} {\bibfnamefont {E.}~\bibnamefont {Emmanouilidou}}, \bibinfo {author} {\bibfnamefont {G.}~\bibnamefont {Yin}}, \bibinfo {author} {\bibfnamefont {S.~K.}\ \bibnamefont {Chong}}, \bibinfo {author} {\bibfnamefont {L.}~\bibnamefont {Tai}}, \bibinfo {author} {\bibfnamefont {N.}~\bibnamefont {Ni}},\ and\ \bibinfo {author} {\bibfnamefont {K.~L.}\ \bibnamefont {Wang}},\ }\bibfield  {title} {\bibinfo {title} {Probing the mesoscopic size limit of quantum anomalous {Hall} insulators},\ }\href {https://doi.org/10.1038/s41467-022-31105-w} {\bibfield  {journal} {\bibinfo  {journal} {Nature Communications}\ }\textbf {\bibinfo {volume} {13}},\ \bibinfo {pages} {4246} (\bibinfo {year} {2022})}\BibitemShut {NoStop}%
\bibitem [{\citenamefont {Long}\ \emph {et~al.}(2008)\citenamefont {Long}, \citenamefont {Sun},\ and\ \citenamefont {Wang}}]{long_disorder-induced_2008}%
  \BibitemOpen
  \bibfield  {author} {\bibinfo {author} {\bibfnamefont {W.}~\bibnamefont {Long}}, \bibinfo {author} {\bibfnamefont {Q.-f.}\ \bibnamefont {Sun}},\ and\ \bibinfo {author} {\bibfnamefont {J.}~\bibnamefont {Wang}},\ }\bibfield  {title} {\bibinfo {title} {Disorder-{Induced} {Enhancement} of {Transport} through {Graphene} p{\textbackslash}mathrm{\textbackslash}text{\textbackslash}ensuremath-n {Junctions}},\ }\href {https://doi.org/10.1103/PhysRevLett.101.166806} {\bibfield  {journal} {\bibinfo  {journal} {Phys. Rev. Lett.}\ }\textbf {\bibinfo {volume} {101}},\ \bibinfo {pages} {166806} (\bibinfo {year} {2008})}\BibitemShut {NoStop}%
\bibitem [{\citenamefont {Liu}\ \emph {et~al.}(2024)\citenamefont {Liu}, \citenamefont {Mao},\ and\ \citenamefont {Sun}}]{liu_four-terminal_2024}%
  \BibitemOpen
  \bibfield  {author} {\bibinfo {author} {\bibfnamefont {P.-Y.}\ \bibnamefont {Liu}}, \bibinfo {author} {\bibfnamefont {Y.}~\bibnamefont {Mao}},\ and\ \bibinfo {author} {\bibfnamefont {Q.-F.}\ \bibnamefont {Sun}},\ }\bibfield  {title} {\bibinfo {title} {Four-terminal graphene-superconductor thermal switch controlled by the superconducting phase difference},\ }\href {https://doi.org/10.1103/PhysRevApplied.21.024001} {\bibfield  {journal} {\bibinfo  {journal} {Phys. Rev. Appl.}\ }\textbf {\bibinfo {volume} {21}},\ \bibinfo {pages} {024001} (\bibinfo {year} {2024})}\BibitemShut {NoStop}%
\end{thebibliography}%

\end{document}